# Understanding Dynamics in Coarse-Grained Models: V. Extension of Coarse-Grained Dynamics Theory to Non-Hard Sphere Systems

Jaehyeok Jin[1,2] and Gregory A. Voth[1,*]

[1] Department of Chemistry, Chicago Center for Theoretical Chemistry, Institute for Biophysical Dynamics, and James Franck Institute, The University of Chicago, Chicago, IL 60637, USA

[2] Department of Chemistry, Columbia University, New York, NY 10027, USA

* Corresponding author: gavoth@uchicago.edu

**Abstract**
Coarse-grained (CG) modeling has gained significant attention in recent years due to its wide applicability in enhancing the spatiotemporal scales of molecular simulations. While CG simulations, often performed with Hamiltonian mechanics, faithfully recapitulate structural correlations at equilibrium, they lead to ambiguously accelerated dynamics. In the first paper of this series [J. Chem. Phys. **158**, 034103 (2023)], we proposed the excess entropy scaling relationship to understand the CG dynamics. Then, in the following companion paper [J. Chem. Phys. **158**, 034104 (2023)], we developed a theory to map the CG system into a dynamically-consistent hard sphere system to analytically derive an expression for fast CG dynamics. However, many chemical and physical systems do not exhibit hard sphere-like behavior, limiting the extensibility of the developed theory. In this paper, we aim to generalize the theory to the non-hard sphere system based on the Weeks-Chandler-Andersen perturbation theory. Since non-hard sphere-like CG interactions affect the excess entropy term as it deviates from the hard sphere description, we explicitly account for the extra entropy to correct the non-hard sphere nature of the system. This approach is demonstrated for two different types of interactions seen in liquids, and we further provide a generalized description for any CG models using the generalized Gaussian CG models using Gaussian basis sets. Altogether, this work allows for extending the range and applicability of the hard sphere CG dynamics theory to a myriad of CG liquids.

## I. Introduction

Classical computer simulations at the atomistic, or fine-grained (FG), resolution have opened a new era of understanding complex chemical and physical phenomena in equilibrium.[1-7] Unfortunately, direct atomistic molecular dynamics (MD) simulations are still computationally infeasible for accessing the relevant spatiotemporal scales pertinent to large-scale systems and processes, e.g., biomolecules and materials. In order to resolve this limitation, coarse-grained (CG) models as reduced low-dimensional models have been proposed by integrating out the unnecessary or slow degrees of freedom,[8-16] which allows for exploring much longer time scales and larger systems at affordable computational costs.[17-21]

Among different strategies for constructing such CG models,[8-16, 22-27] bottom-up CG models are constructed based on more detailed, often all-atomistic, models. By designing the effective CG model interaction as a many-body potential of mean force (PMF),[15, 28, 29] the equilibrium structural correlations observed at the FG resolution are still reproduced in these bottom-up CG models. For example, using pairwise basis sets for the CG PMF, relative entropy minimization (REM)[30-32] is in principle guaranteed to capture the two-body correlations,[33] and the multi-scale coarse-graining (MS-CG) methodology[34-38] is also known to capture up to three-body correlations.[39] Despite its capability in reproducing important structural correlations, the resultant CG dynamics under the CG simulation still have areas for improvement.[40]

Conventional bottom-up CG simulations are typically propagated using the Hamiltonian mechanics, such that the many-body CG PMF becomes the system Hamiltonian.[41] Hence, in the absence of correct fluctuational and frictional forces during a CG simulation, the overall dynamics are expected to be accelerated, resulting in an overestimation of the diffusion coefficients.[40, 42-46] It should be noted that some approaches such as time-rescaling aim to rescale the ambiguous accelerated CG time scales.[44, 47-49] However, the speed-up factor originates from the complex many-body effects embedded in the missing frictional forces at the CG resolution, and it is not straightforward to elucidate the underlying physical nature of the speed-up factor. In principle, a rigorous theoretical approach to addressing the dynamics in such reduced models can be built upon the Mori-Zwanzig formalism.[50-53] Nevertheless, due to the complex nature of frictional kernels that are correlated with both temporal and configurational variables, applications to more complex systems, e.g., heterogeneous system, significantly suffer from numerical limitations.[43, 54]

As an alternative approach, in the current series of papers, we introduce the excess entropy scaling relationship that has received steadily growing attention.[55] From the relatively simple point particle simulations, such as the Lennard-Jones system, Rosenfeld originally proposed that dynamical properties such as viscosity and diffusion coefficients are exponential functions of the molar excess entropy of the system.[56-58] We are particularly interested in the self-diffusion coefficients $D$ that satisfy the following relationship

$$D^* = D_0 \exp(\alpha s_{ex}) \,. \tag{1}$$

In Eq. (1), $D^*$ is a dimensionless, macroscopically scaled diffusion coefficient, derived from $D$, expressed as

$$D^* = D\frac{\rho^{\frac{1}{3}}}{\left(\frac{k_B T}{m}\right)^{\frac{1}{2}}}, \tag{2}$$

and $s_{ex}$ is the molar excess entropy of the system. Note that the excess thermodynamic quantity is a measure of how much the system deviates from the ideal gas at the thermodynamic state points (given by number density $\rho$ and temperature $T$),

$$s_{ex} = \frac{S_{ex}}{Nk_B} = \frac{1}{Nk_B}\left(S(\rho, T) - S_{\mathrm{id}}(\rho, T)\right), \tag{3}$$

where $S(\rho, T)$ and $S_{\mathrm{id}}(\rho, T)$ correspond to the system entropy and the ideal gas entropy, respectively. Since this scaling relationship has been widely applied to different systems ranging from molecular liquids[56] to complex systems such as ionic liquids[59] and biomolecules,[60] the previous papers of the series[61-64] focused on developing a full FG–CG dynamic correspondence on the basis of this scaling relationship.[65, 66] Hereafter, we denote the previous papers of this series as Papers I,[61] II,[62] III,[63] and IV.[64]

The major caveat of the excess entropy scaling relationship is that the relationship itself cannot be derived from rigorous statistical mechanical theories,[55] and it has been considered as a semi-quantitative model due to this approximative nature.[67] This is because, even though $D^*$ and $s_{ex}$ in Eq. (1) correspond to physical quantities, $D_0$ does not have any clear physical correspondence. Despite partial success in understanding this relationship from microscopic physics under limited conditions[68-72] and in simple point-particle systems via isomorph theory,[73, 74] the phenomenological nature of $D_0$ makes it a major challenge to clarifying the CG dynamics. Namely, understanding $D_0$ is crucial for elucidating the scaling relationship. Since entropic contributions from the diffusion coefficient $D^*$ are separated as $\exp(\alpha s_{ex})$ terms, we further denote $D_0$ as the "entropy-free" diffusion coefficient of the system.

To overcome the aforementioned phenomenological limitation, we introduced in Paper II[62] an additional layer of coarsening the CG system into its dynamically-consistent hard sphere system, where the thermodynamic and dynamical properties can be analytically expressed as a function of the packing fraction $\eta$:

$$\eta = \frac{\pi}{6}\sigma^3\rho. \tag{4}$$

Since the bulk number density $\rho$ is a fixed value, we developed a statistical mechanical theory[75] for obtaining the effective hard sphere diameter $\sigma$ of the CG system. Along with the conventional Barker-Henderson[76, 77] and Weeks-Chandler-Andersen[78-80] perturbation theories, we were able to demonstrate that the CG model for water[81, 82] can be faithfully represented by the hard sphere description, yielding an analytical expression for the entropy-free diffusion coefficient $D_0$ for water.[62] The basic idea of this "hard sphere mapping theory" is to extract the hard sphere characteristics by looking at the pairwise CG interactions (akin to Barker-Henderson[76, 77]) or the long wavelength density fluctuations (akin to Weeks-Chandler-Andersen[78-80] and Mirigian-Schweizer[83-85]). Despite the success of this mapping theory, there is a limitation to this theory that prevents its applicability to various chemical systems. Since the excess entropy and diffusion coefficient can be approximately reduced as a function of only $\eta$ under the hard sphere system, this theory becomes no longer valid for the many other systems that do not exhibit hard sphere characteristics.

The basic underlying assumption of Paper II was from the classical perturbation theory of liquids that hard sphere-like repulsions at short distances primarily determine the structure of liquids, and long-range attractions act as a perturbation to the reference repulsion.[62] However, for any CG models that exhibit non-trivial interactions in the short-range regime, the overall CG interaction is likely to deviate from the hard sphere expectation. A good theoretical example to note is the Pratt-Chandler theory, where the non-hard sphere interactions are considered to treat the hydrophobic effect.[86] In a much more complicated manner, biomolecules with various interactions are expected to significantly deviate from this classical hard sphere model. For example, electrostatic interactions in these complex systems, e.g., electric double layer,[87, 88] have non-hard sphere contributions. In addition to these electrostatic origins, various sources for the "softness" emerge due to the steric effect and temperature-dependent behavior in many different systems, e.g., colloidal systems.[89] Also, self-assembly behaviors of polymers and proteins are way beyond the pure hard sphere phenomena, as can be discerned from the corresponding experiments with less structured scattering intensities.[90-92] Hence, to extend the applicability of the previously developed framework to the CG dynamics, in this paper, we seek to generalize the scope of the CG dynamics theory to non-hard sphere systems.

When the system consists of non-repulsive short-range interactions, our approach to employ the Barker-Henderson perturbation theory[76, 77] to infer the effective hard sphere diameter is no longer valid since the overall interaction does not decay to zero in the short-range regime. However, the local structural orderings at these short distances can still be attributed to the effective hard sphere repulsion. Therefore, even for non-hard sphere systems, an effective hard sphere repulsion can still be effectively considered using the Weeks-Chandler-Andersen theory that is based on the sign of the forces.[78-80] Nevertheless, this will introduce a non-trivial correction factor to $D_0$ to address the transformation from a hard sphere system to a non-hard sphere system. By carefully designing the non-hard sphere system and accounting for the correction factor, we expect that the developed framework will be applicable to arbitrary systems. Even though an effective hard sphere mapping of the non-hard sphere system is not a new idea, as the perturbation treatment using a soft-sphere reference system[93] has been widely used, no other work in literature so far has accounted for the physical nature underlying the excess entropy scaling relationship of non-hard sphere systems to the best of our knowledge.

In this paper, we provide an expanded theory and results for this non-hard sphere correction scheme. In Sec. II, we briefly review the excess entropy scaling and the hard sphere CG dynamics theory. After examining the limitations of the previous approach, we present the non-hard sphere CG dynamics theory. In Sec. III, we apply the non-hard sphere CG dynamics theory to two different types of liquids. As an extension of our recent work,[94] we then generalize the non-hard sphere correction scheme for any CG system by representing a CG interaction using Gaussian basis sets. The work presented allows one to determine the correction factor for any CG system, extending the CG dynamics theory to a wide range of chemical and physical systems of potential interest.

## II. Hard Sphere CG Dynamics Theory

## A. Excess Entropy Scaling

Rosenfeld's scaling relationship[56-58] introduced in Eq. (1) can be further reorganized to describe the reduced diffusion coefficient of the CG system

$$D_{\mathrm{CG}}^{*} = D_{0}^{\mathrm{CG}} \exp(\alpha s_{ex}^{\mathrm{CG}}), \tag{5}$$

where the molar CG entropy $s^{\mathrm{CG}}$ is always smaller than the molar FG entropy $s^{\mathrm{FG}}$, and thus the excess entropy difference is known as the mapping or "missing" entropy.[95, 96] Especially, for the single-site CG mapping where the molecule is mapped into a single site (or "bead") at the center-of-mass position, the translational motions are only observed in the CG system. References 95, 96 provide a detailed explanation of the missing rotation and vibrational contributions in terms of the CG entropy, and Paper III extensively discusses the role of these missing motions to understand the FG dynamics.[61] Lastly, Paper IV generalizes the findings from Paper III by utilizing the hydrodynamic perspective to the missing diffusion described by the Stokes-Einstein and Stokes-Einstein-Debye relationships.[64]

It is known that the translational entropy of the ideal gas follows the Sackur-Tetrode equation[97, 98]

$$s_{\mathrm{trn}}^{(\mathrm{id})} = \frac{S_{\mathrm{trn}}^{(\mathrm{id})}}{Nk_B} = -\ln\left(\frac{h^2}{2\pi m k_B T}\right)^{\frac{3}{2}} - \ln\left(\frac{N}{V}\right) + \frac{5}{2}, \tag{6}$$

where $k_B$ and $h$ are the Boltzmann and Planck constants, respectively, and $N/V$ gives the number density of the system, $\rho$. By combining Eq. (3) with Eq. (6), we arrive at the excess entropy scaling relationship for CG systems

$$s_{ex}^{\mathrm{CG}} = s^{\mathrm{CG}} - \left[\frac{5}{2} - \ln\left(\frac{h^2}{2\pi m k_B T}\right)^{\frac{3}{2}} - \ln\left(\frac{N}{V}\right)\right]. \tag{7}$$

The molar CG entropy $s^{\mathrm{CG}}$ can be determined using numerous methods available in literature.[99, 100] In this work, we followed the same protocol from the companion papers by employing the two-phase thermodynamic (2PT) method developed by Goddard and co-workers.[99-101] In brief, the 2PT method calculates the thermodynamic property of the system by constructing the canonical partition function as a combination of solid-like and gas-like components. Theoretical discussions of the 2PT methodology are detailed in Refs. 99-101, and a thorough review of the applicability of the 2PT method to the excess entropy scaling formalism is found in Ref. 61. Numerical validations of excess entropy scaling for paradigmatic systems (e.g., two-dimensional or soft rods) and complex atomistic systems, as presented in Refs. 102-104, further support extending its applicability to bottom-up CG systems.

## B. Hard Sphere Mapping Theory

In Paper II, among several hard sphere mapping approaches,[105] we proposed two distinct approaches to effectively map the CG system to the corresponding hard sphere systems to determine $D_0^{\mathrm{CG}}$ from Eq. (5).[62] The first choice is the Barker-Henderson diameter $\sigma_{\mathrm{BH}}$ that is derived directly from the Barker-Henderson perturbation theory[76, 77]

$$\sigma_{\mathrm{BH}} = \int_0^{R_0} \left[1 - \exp\left(-\beta U(R)\right)\right] dR, \tag{8}$$

where $R_0$ is the smallest distance at which the short-range repulsive interaction vanishes, i.e., $U(R_0) = 0$. In other words, $\sigma_{\mathrm{BH}}$ directly maps the repulsive characteristic seen from the CG interaction to a hard sphere diameter, and it can be obtained straightforwardly from the pairwise CG interaction $U(R)$.

Alternatively, the second method named "fluctuation matching" considers the effect of long-range density fluctuations from the CG dynamics by matching the density fluctuations[83-85] that are encoded in the isothermal compressibility $\kappa_T = -(\partial V/\partial P)_T/V$ via[106]

$$S(k=0) = \rho k_B T \kappa_T, \tag{9}$$

where $S(k=0)$ denotes the structure factor at the zero wave number limit. In this limit, the Fourier transformed structure factor is expressed as a simplified function of the pair correlation function, or radial distribution function (RDF), $g(R)$,

$$S(k=0) = 1 + 4\pi\rho \int_0^\infty dr\, R^2[g(R)-1]\,. \tag{10}$$

By matching the $S(k=0)_{\mathrm{CG}}$ to $S(k=0)_{\mathrm{HS}}$, we arrive at

$$1 + 4\pi\rho \int_0^\infty dr\, R^2[g(R)-1] = \rho k_B T \kappa_T. \tag{11}$$

In Eq. (11), $S(k=0)_{\mathrm{CG}}$ can be computed from the CG simulation (left-hand side). Since the right-hand side of Eq. (11) depends solely on the packing fraction, the gist of fluctuation matching lies in determining a packing fraction $\eta$ that recapitulates the density fluctuations observed in the CG system. Additionally, in Eq. (11), we accounted for the finite-size effect, as thoroughly discussed in Ref. 62.

### C. Entropy-Free Diffusion Term

By mapping the CG system to an effective hard sphere system following Eqs. (8) or (11), one can then derive an analytical expression of $D_0^{\mathrm{CG}}$ for different types of equations of state (EOSs) using the Enskog kinetic theory.[107, 108] In Paper II,[62] we considered three EOSs that have been generally employed in literature.[106, 109-111] It should be noted that the EOS is often expressed as a compressibility factor $\mathbb{Z}$, which is different from the isothermal compressibility in Eq. (9):

$$\mathbb{Z} = \frac{P}{\rho k_B T}. \tag{12}$$

First, the Percus-Yevick EOS is written as[112]

$$\mathbb{Z}_{\mathrm{PY}} = \frac{(1+\eta+\eta^2)}{(1-\eta)^3}, \tag{13}$$

and the second EOS is suggested by Carnahan-Starling[113]

$$\mathbb{Z}_{\mathrm{CS}} = \frac{1+\eta+\eta^2-\eta^3}{(1-\eta)^3}. \tag{14}$$

The last EOS is a slightly improved Carnahan-Starling EOS designed by Kolafa[114, 115]

$$\mathbb{Z}_{\mathrm{CSK}} = \frac{1+\eta+\eta^2-\frac{2}{3}(\eta^3+\eta^4)}{(1-\eta)^3}. \tag{15}$$

Using these different EOSs, in Paper II,[62] we derived the $D_0^{\mathrm{CG}}$ for each of these EOS as

$$D_{0,\mathrm{PY}}^{\mathrm{HS}} \approx \frac{6^{\frac{1}{3}}}{4}\pi^{\frac{1}{6}}\eta^{\frac{1}{3}}\frac{(1-\eta)^2}{\eta(\eta^2-2\eta+4)}\exp\left[\frac{3(2\eta-\eta^2)}{2(1-\eta)^2}\right], \tag{16a}$$

$$D_{0,\mathrm{CS}}^{\mathrm{HS}} \approx \frac{\pi^{\frac{1}{6}}}{48} \cdot 6^{\frac{4}{3}} \cdot \frac{(1-\eta)^3}{\eta^{\frac{2}{3}}(2-\eta)} \exp\left[\frac{(4\eta-3\eta^2)}{(1-\eta)^2}\right], \tag{16b}$$

$$D_{0,\mathrm{CSK}}^{\mathrm{HS}} \approx \frac{\pi^{\frac{1}{6}}}{96} \cdot 6^{\frac{4}{3}} \cdot \frac{(1-\eta)^{\frac{14}{3}}}{\eta^{\frac{2}{3}}\left(1-\frac{\eta}{2}+\frac{\eta^2}{12}-\frac{\eta^3}{6}\right)} \exp\left[\eta\frac{4\eta^2-33\eta+34}{6(1-\eta)^2}\right]. \tag{16c}$$

We also have demonstrated that Eqs. (16a)–(16c) can be readily applied to the different FG force fields of water. In terms of accuracy in reproducing diffusion coefficients, both the Barker-Henderson and fluctuation matching methods provide reasonable estimates, with errors of 16.0% and 15.1%, respectively.[62]

## III. Theory

### A. Bottom-Up Coarse-Graining

In principle, bottom-up CG models are designed to reproduce the many-body CG PMFs of the simplified CG system at the reduced resolution.[34-37, 39] Among other CG methodologies including the REM[30-33] and the generalized Yvon-Born-Green methodologies,[116-121] we are particularly concerned with the MS-CG methodologies in this work. A primary goal of MS-CG is to obtain the effective CG forces by minimizing the force residuals $\chi^2[\mathbf{F}]$ with respect to the FG reference $\mathbf{f}_I(\mathbf{r}^n)$[34-38]

$$\chi^2[\mathbf{F}] = \frac{1}{3N}\langle\sum_{I=1}^{N}\left|\mathbf{f}_I(\mathbf{r}^n)-\mathbf{F}_I\left(M_{\mathbf{R}}^N(\mathbf{r}^n)\right)\right|^2\rangle, \tag{17}$$

where $\mathbf{r}^n$ and $\mathbf{R}^N$ denote the FG and CG configurations, respectively. These two configurational variables are linked via the mapping operator $M_{\mathbf{R}}^N(\mathbf{r}^n)$: $\mathbf{r}^n \rightarrow \mathbf{R}^N$. Namely, Eq. (5) tries to match the CG forces $\mathbf{F}_I(M_{\mathbf{R}}^N(\mathbf{r}^n))$ to the FG forces exerted on the CG site *I*, $\mathbf{f}_I(\mathbf{r}^n)$. Thus, the CG force expression is determined by the least square regression for the CG basis sets. In a similar manner as in the companion papers of this series,[61-64] this is done by utilizing the pairwise basis sets

$$\mathbf{F}_I\left(M_{\mathbf{R}}^N(\mathbf{r}^n)\right) = \mathbf{F}_I(\mathbf{R}^N) = \sum_{I\neq J} F\left(R_{IJ}\right). \tag{18}$$

### B. Limitations

Equations (16a)–(16c) suggest that the entropy-free diffusion coefficient of hard spheres is only a function of the packing fraction $\eta$. Since $\eta$ is a bounded quantity, we now examine the numerical bounds of Eqs. (16a)–(16b) to see if there is any limitation of employing the hard sphere mapping theory to the CG system for understanding the entropy-free dynamics. Since the Carnahan-Starling-Kolafa EOS is similar to the Carnahan-Starling EOS, $dD_{0,CSK}^{\mathrm{HS}}/d\eta$ is omitted here for clarity.

By taking a derivative of Eqs. (16a)–(16b) with respect to $\eta$, $dD_0/d\eta$ is written as

$$\frac{dD_{0,\mathrm{PY}}^{\mathrm{HS}}}{d\eta} \approx \frac{\pi^{\frac{1}{6}}}{2\cdot 6^{\frac{2}{3}}} \frac{(-2\eta^5+10\eta^4-17\eta^3+20\eta^2-46\eta+8)}{\eta^{\frac{5}{3}}(\eta-1)(\eta^2-2\eta+4)^2} \exp\left[\frac{3(2\eta-\eta^2)}{2(1-\eta)^2}\right], \tag{19a}$$

$$\frac{dD_{0,\mathrm{CS}}^{\mathrm{HS}}}{d\eta} \approx \frac{\pi^{\frac{1}{6}}}{4\cdot 6^{\frac{2}{3}}}\cdot 6^{\frac{4}{3}}\cdot\frac{4\eta^4-11\eta^3-6\eta^2+23\eta-4}{\eta^{\frac{5}{3}}(\eta-2)^2}\exp\left[\frac{(4\eta-3\eta^2)}{(1-\eta)^2}\right]. \qquad (19b)$$

By examining the values of $-2\eta^5+10\eta^4-17\eta^3+20\eta^2-46\eta+8$ from $dD_{0,PY}^{HS}/d\eta$ and $4\eta^4-11\eta^3-6\eta^2+23\eta-4$ from $dD_{0,\mathrm{CS}}^{\mathrm{HS}}/d\eta$ in $\eta\in[0,1]$, it is clear that both $D_0^{\mathrm{CG}}$ terms have only one minimum in [0,1]. Especially, $-2\eta^5+10\eta^4-17\eta^3+20\eta^2-46\eta+8$ monotonically decreases in [0,1] from 8 to -27, indicating that $D_{0,\mathrm{PY}}^{\mathrm{HS}}$ diverges as $\eta\to 0$ and $\eta\to 1$. For $D_{0,\mathrm{CS}}^{\mathrm{HS}}$, the derivative $dD_{0,\mathrm{CS}}^{\mathrm{HS}}/d\eta$ monotonically increases from zero to $\eta=0.8139$. Thus, in physically relevant $\eta$ regimes, $dD_{0,CS}^{\mathrm{HS}}/d\eta$ has one minimum value and increases as $\eta$ deviates from the minimum value.

Figure 1 depicts the behavior of $D_0^{\mathrm{HS}}$ as $\eta$ varies between 0 and 0.6. We plot the packing fraction up to 0.6 since the theoretically maximum packing fraction value is known to be $\eta_{\mathrm{max}}=0.545$. As shown in the case of water,[62] different EOSs do not significantly affect the $D_0^{\mathrm{HS}}$ values. Therefore, without loss of generality, we choose the simplest Percus-Yevick EOS for describing the hard sphere system hereafter. It should be noted that other EOSs can be also used regardless of its specific form. As an example, we will conduct the same procedure using the Carnahan-Starling EOS and provide the results in Appendix A.

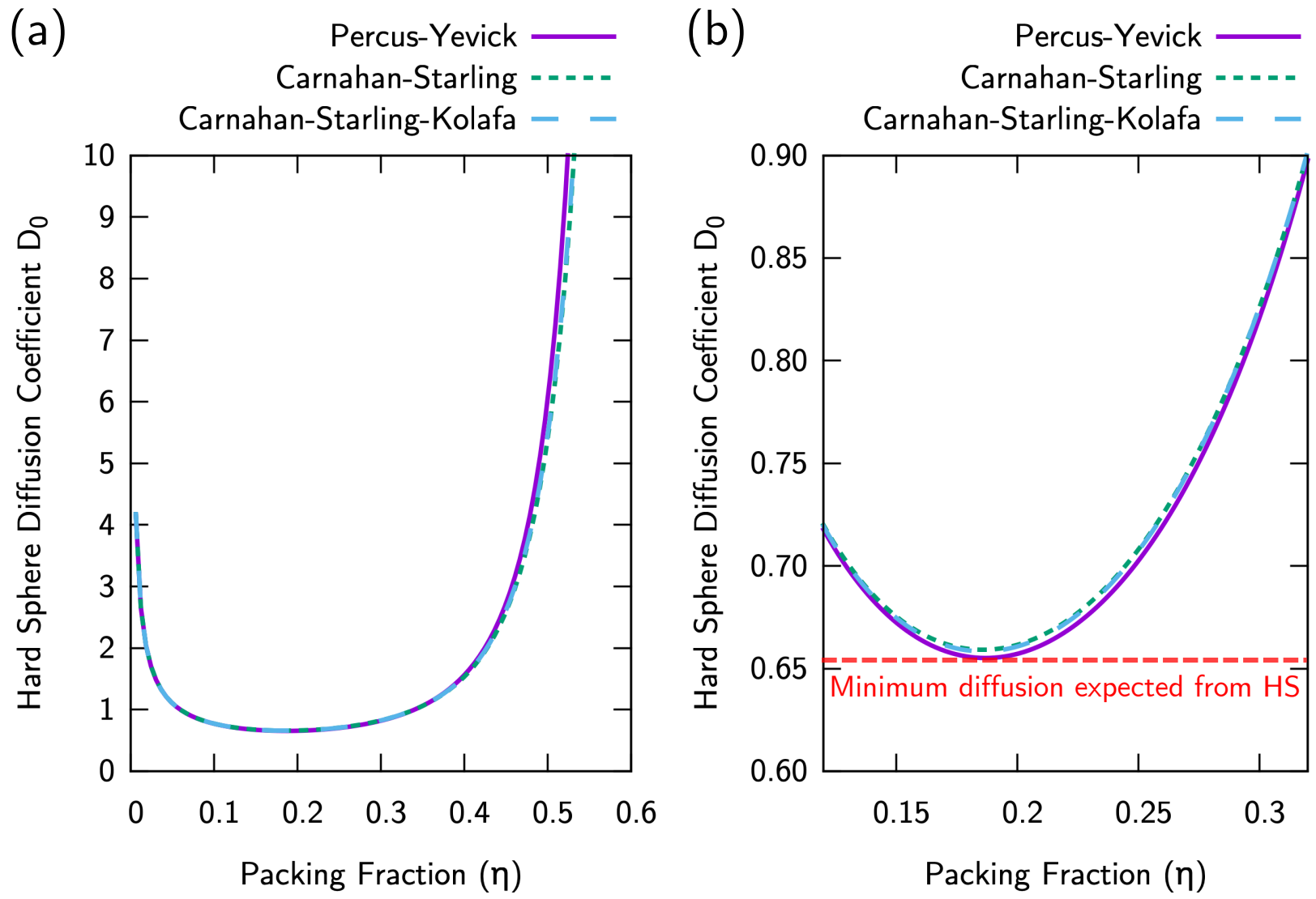

**Figure 1.** Lower bound analysis of the entropy-free diffusion coefficient $D_0$ using different hard sphere EOSs: Percus-Yevick (purple line), Carnahan-Starling (green dashes), and Carnahan-Starling-Kolafa (blue dashes). (a) Estimated $D_0$ from $\eta=0$ to 0.6. (b) Magnified plot near $\eta\approx 0.2$, where the derivative of $D_0$ indicates that there is a minimum $D_0$ value (red dashes) expected from the hard sphere CG dynamics theory.

Interestingly, Fig. 1 suggests that there is a minimum diffusion phenomenon at $\eta_{\mathrm{min}}$ exerted by hard sphere collisions. For systems with packing densities lower than $\eta_{\mathrm{min}}$, the diffusion is enhanced as it is entropically favored. In the other limit, where $\eta>\eta_{\mathrm{min}}$, hard spheres experience more frequent collisions that will eventually enhance the collective diffusion. This analysis immediately shows that the hard sphere model is not able to reproduce slow diffusion processes $D_0^{\mathrm{CG}} < 0.65$. However, there is no guarantee that any other CG systems will satisfy this criterion, implying that a hard sphere description will no longer be valid for these systems.

**Table 1.** Lower bound of the hard sphere entropy-free diffusion coefficient $D_{0,\mathrm{min}}(\eta_{\mathrm{min}})$ estimated from the different EOSs: Percus-Yevick, Carnahan-Starling, and Carnahan-Starling-Kolafa at its minimum packing fraction $\eta_{\mathrm{min}}$.

| | **Equations of State** | | |
|---|---|---|---|
| | **Percus-Yevick** | **Carnahan-Starling** | **Carnahan-Starling-Kolafa** |
| $\eta_{\mathrm{min}}$ | 0.1869 | 0.1858 | 0.1859 |
| $D_{0,\mathrm{min}}(\eta_{\mathrm{min}})$ | 0.655 | 0.659 | 0.658 |

### C. Non-Hard Sphere System

Even though the Barker-Henderson diameter and fluctuation matching can be utilized to effectively extract the hard sphere repulsion, Eq. (16) breaks down for certain systems that are slightly perturbed from the hard sphere description. As a representative example of the non-hard sphere system, the left panel of Fig. 2, taken from Paper II,[62] shows the CG methanol interaction at 300K. As witnessed from the interaction profile, the effective repulsion from the hard sphere description is present in the inner-core region $R < 3$Å, but there is an additional repulsion at 3–5Å. This additional repulsive interaction that gives rise to the first coordination shell near 3–5Å prevents the Barker-Henderson integrand $[1 - \exp(-\beta U(R))]$ from decaying to zero, resulting in much larger effective hard sphere diameter values than expected.

This breakdown will also occur in other complex CG models if their interactions have additional repulsions or attractions located at the pair distances beyond the inner-core region, which significantly affect the convergence of the $[1 - \exp(-\beta U(R))]$ term. Here, we opt to describe a breakdown scenario of the hard sphere assumption using the Barker-Henderson approach under the established framework from Paper II, and we will discuss the feasibility of fluctuation matching in later sections.

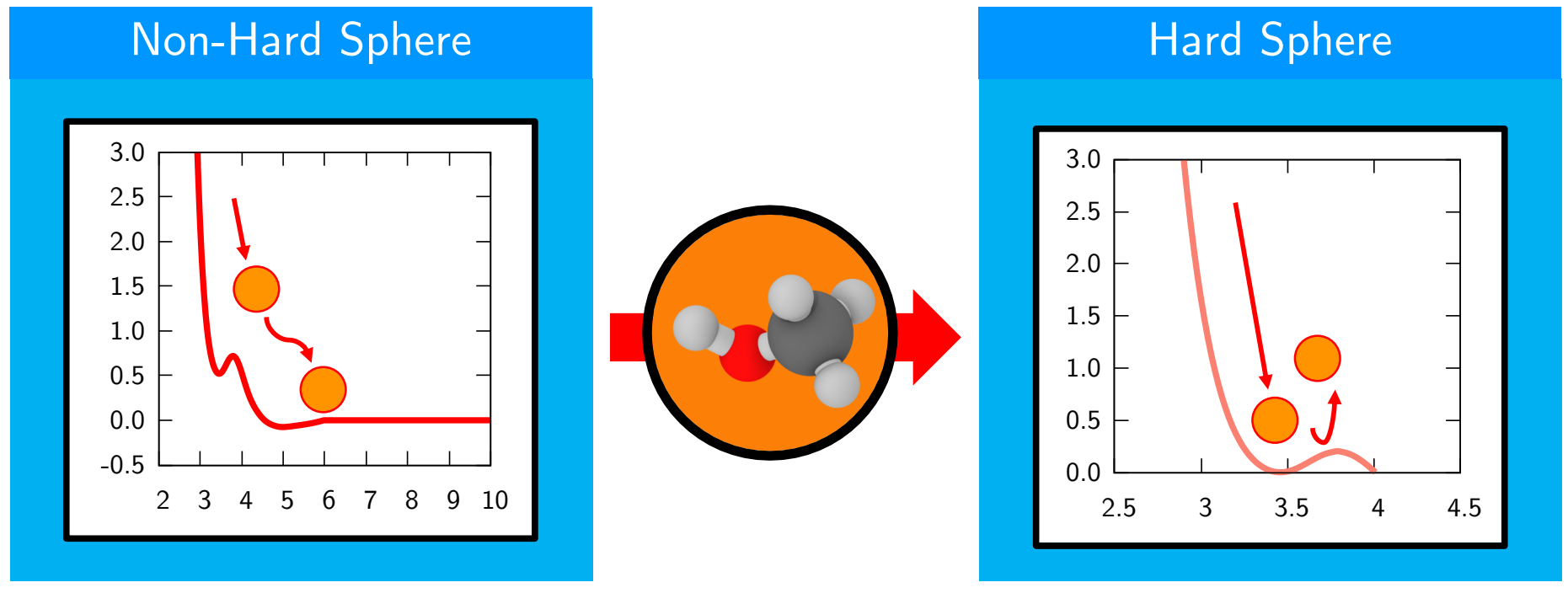

**Figure 2.** The non-hard sphere nature of the CG interaction seen from the single-site CG methanol model. In the left panel, the hard sphere character contributes to a local ordering near 3.5Å, but the overall interaction slowly decays due to the additional repulsion after 3.5Å. From the Weeks-Chandler-Andersen theory, this effective hard sphere repulsion at the inner-core can be adjusted (*hybrid* Barker-Henderson[122, 123]), as shown in the right panel.

As pointed out by Weeks, Chandler, and Andersen,[78-80] the hard sphere nature can be separated on the basis of the sign of forces, not the sign of potentials (Barker-Henderson approach[76, 77]) known as the Weeks separation. To incorporate the Weeks-Chandler-Andersen theory to the Barker-Henderson approach, we adjust the non-hard sphere-like interaction $U(R)$ by shifting the potential value $U(R_{\text{min}})$ at the smallest local minimum $R_{\text{min}}$, where the pair force is zero, to $U = 0$,

$$U^{\text{HS}}(R) = \omega(R)[U(R) - U(R_{\text{min}})], \tag{20}$$

where $\omega(R) \in [0,1]$ is the radial kernel following the sigmoid form, which is approximately 1 for $R \leq R_{\text{min}}$ and smoothly decays to zero for $R_{\text{min}} < R < R_{\text{cutoff}}$. Here, $R_{\text{cutoff}}$ is chosen to be half the system size to ensure that $U^{\text{HS}}(R)$ properly captures the effective forces responsible for hard sphere behavior while decaying to zero at large distances.

In the case of methanol, $R_{\text{min}}$ is 3.46Å, and the shifted CG interaction is shown in the right panel of Fig. 2. By employing Eq. (20), the effective hard sphere regime that is consistent with the Weeks-Chandler-Andersen criteria can be readily obtained via the Barker-Henderson approach. Applying Eq. (20) to Eq. (8), the corrected hard sphere diameter $\sigma'_{\text{BH}}$ determined from

$$\sigma'_{\text{BH}} = \int_0^{R'_0} [1 - \exp(-\beta\omega(R)\{U(R) - U(R_{\text{min}})\})]dR\,, \tag{21}$$

is expected to capture the correct hard sphere behavior. Here, $R'_0$ is the minimum distance where $U^{\text{HS}}(R'_0) \approx U(R'_0) - U(R_{\text{min}}) = 0$, and thus $R'_0 = R_{\text{min}}$ holds once Eq. (20) is applied. However, we note that, in general, the minimum zero-force distance and minimum zero-potential distance are not necessarily equivalent (i.e., $R_{\text{min}} \neq R_0$) for complex CG potentials. This discrepancy can lead to inaccuracies in describing the effective hard sphere behavior. This separation scheme shares a similar principle with the *hybrid* Barker-Henderson approximation in the literature,[122, 123] because Eqs. (20) and (21) incorporate the Weeks-Chandler-Andersen separation into the Barker-Henderson perturbative treatment. We note that the conventional hybrid Barker-Henderson approximation is mainly applied to the Lennard-Jones system to construct the perturbative interaction as follows: the reference potential $U^{\text{HS}}(R)$ is defined as $U(R) + \epsilon$ for $R < R_0$ and 0 otherwise, while the perturbation $U^{(1)}(R)$ is given by $-\epsilon$ for $R < R_0$ and $U(R)$ otherwise. Since the attraction strength $\epsilon$ is positive in Lennard-Jones interactions, this conventional treatment has primarily been established for attractive perturbations in relatively simple models. Thus, generalization to $\epsilon < 0$ is necessary for addressing non-hard sphere liquid CG interactions with additional repulsion, as recently demonstrated in Ref. 124.

This non-trivial adjustment of the CG interaction will unambiguously affect the structural correlation of the shifted CG system and the excess entropy value as well. Alternatively, $\sigma'_{\text{BH}}$ for the non-hard sphere system can be also obtained from the compressibility relationship that fits the packing fraction from the calculated compressibility values at different temperatures.[125] We will discuss these two approaches (Barker-Henderson theory and the compressibility relationship) in the next subsection, Sec. III E.

### D. Non-Hard Sphere System: Correction Factor

In Sec. II C, we derived the expression for $D_0^{\text{HS}}$, which describes the entropy-free diffusion coefficient derived from the hard sphere EOSs. For example, the Percus-Yevick EOS gives

$$D_0^{\mathrm{HS}} \approx \frac{6^{\frac{1}{3}}}{4}\pi^{\frac{1}{6}}\eta^{\frac{1}{3}}\frac{(1-\eta)^2}{\eta(\eta^2-2\eta+4)}\exp\left[\frac{3(2\eta-\eta^2)}{2(1-\eta)^2}\right]. \tag{22}$$

By shifting the CG interaction, one can obtain a correct $\eta$ value from $\sigma_{\mathrm{BH}}$, but the excess entropy value of the shifted system will vary at the same time. In other words, while deriving Eq. (22), the excess entropy part was integrated using

$$s_{ex}^{\mathrm{HS}} = -\int_0^{\eta}\frac{\eta'^2-2\eta+4}{(1-\eta')^3}d\eta' = \ln(1-\eta)-\frac{3}{2}\left[\frac{1}{(1-\eta)^2}-1\right]. \tag{23}$$

The excess entropy part can then be isolated from Eq. (22)

$$D_0^{\mathrm{HS}} \approx \frac{6^{\frac{1}{3}}}{4}\pi^{\frac{1}{6}}\eta^{\frac{1}{3}}\frac{(1-\eta)^3}{\eta(\eta^2-2\eta+4)}\exp(s_{ex}). \tag{24}$$

For the non-hard sphere system, Eq. (24) cannot be reduced to Eq. (22) since the excess entropy of the system $s_{ex}$ is not equivalent to $s_{ex}^{\mathrm{HS}}$, indicating that $D_0^{\mathrm{HS}}$ from the hard sphere description is not equivalent to $D_0$: $D_0^{\mathrm{HS}} \neq D_0$. Therefore, one should correct this entropy change by introducing a correction factor

$$f = \frac{\exp\left(s_{ex}^{\mathrm{unshifted}}\right)}{\exp\left(s_{ex}^{\mathrm{shifted}}\right)}. \tag{25}$$

From the definition of excess entropy, $f$ simplifies to $\exp\left(s_{ex}^{\mathrm{unshifted}} - s_{ex}^{\mathrm{shifted}}\right)$ and quantifies the total entropy difference between the unshifted and shifted system under perturbation, as given by Eq. (20). Similar to fugacity and activity, which describe molar free energy corrections relative to ideal states, this entropy-based factor corrects for the diffusion coefficients of the original unshifted system. For example, in a rugged energy landscape framework,[126-128] the diffusion coefficient depends on the mean first passage time, which scales exponentially with entropy differences. Namely, by introducing the correction factor $f$ on both left- and right-hand sides, the correct $D_0$ expression for general non-hard sphere systems can be formulated as

$$D_0^{\mathrm{HS}}f \approx \frac{6^{\frac{1}{3}}}{4}\pi^{\frac{1}{6}}\eta^{\frac{1}{3}}\frac{(1-\eta)^3}{\eta(\eta^2-2\eta+4)}\exp(-s_{ex})\,f = \frac{6^{\frac{1}{3}}}{4}\pi^{\frac{1}{6}}\eta^{\frac{1}{3}}\frac{(1-\eta)^3}{\eta(\eta^2-2\eta+4)}\exp\left(-s_{ex}^{\mathrm{shifted}}\right). \tag{26}$$

Since the right-hand side of Eq. (26) correctly contains the excess entropy contribution from the non-hard sphere system, the left-hand side of Eq. (26) is equal to $D_0$, i.e., $D_0 \coloneqq D_0^{\mathrm{HS}}f$. With this in mind, we now discuss how to calculate the correction factor $f$ from the non-hard sphere system. It has been shown that the excess entropy term at the FG resolution contains various contributions from *n*-body multiparticle correlation functions:

$$S_{ex} = \sum_{n\geq 2} S^{(n)} \tag{27}$$

However, at the CG resolution, it is reasonable to assume that the two-body contribution is dominant.[129-132] From Wallace's seminal work,[133-136] the $S^{(2)}$ term is written as a function from the pair distribution function $g^{(2)}(\mathbf{R})$

$$S^{(2)} = -2\pi\rho\int_0^{\infty}\left\{g^{(2)}(\mathbf{R})\ln g^{(2)}(\mathbf{R}) - \left[g^{(2)}(\mathbf{R})-1\right]\right\}\mathbf{R}^2 d\mathbf{R}, \tag{28}$$

where $\mathbf{R}$ is a vector that includes not only the pair distances $R = |\mathbf{R}|$ but also the relative orientations between the pairs. Thus, an exact determination of Eq. (28) at the FG resolution requires an effective factorization of translational motions from the relative orientational motions,[137, 138] as described in Paper I.[61] Nevertheless, for the single-site CG model, there is no effective orientation between particle pairs. By leaving out the orientation vectors, Eq. (28) becomes

$$S^{(2)} = -2\pi\rho \int_0^\infty \left\{ g^{(2)}(R) \ln g^{(2)}(R) - \left[ g^{(2)}(R) - 1 \right] \right\} R^2 dR \,. \tag{29}$$

In principle, combining Eqs. (25) and (29), the correction factor is expressed as

$$f = \frac{\exp\left( -\frac{2\pi\rho}{Nk} \int_0^\infty \left\{ g^{(2)}_{\mathrm{unshifted}}(R) \ln g^{(2)}_{\mathrm{unshifted}}(R) - \left[ g^{(2)}_{\mathrm{unshifted}}(R) - 1 \right] \right\} R^2 dR \right)}{\exp\left( -\frac{2\pi\rho}{Nk} \int_0^\infty \left\{ g^{(2)}_{\mathrm{shifted}}(R) \ln g^{(2)}_{\mathrm{shifted}}(R) - \left[ g^{(2)}_{\mathrm{shifted}}(R) - 1 \right] \right\} R^2 dR \right)} . \tag{30}$$

We further make an approximation to estimate $f$ without requiring additional simulations. Namely, although the pair correlation $g^{(2)}(R)$ denotes the probability distribution of the pair sites in equilibrium, it effectively includes not only the two-body but also the three-body and higher-order correlations. For hard spheres, these higher-order interactions are less important since the primary interest is the collision processes between the CG particles.[129-132] Therefore, one can approximate that the higher-order effects are not affected by adjusting the non-hard sphere CG interactions. This approximation leaves out only the pairwise contributions to the effective excess entropy term such that $g^{(2)}_{\mathrm{eff}}(R) \approx e^{-\beta U(R)}$, giving

$$f \approx \frac{\exp\left( -\frac{2\pi\rho}{Nk} \int_0^\infty \left\{ -\beta e^{-\beta U(R)} U(R) - \left[ e^{-\beta U(R)} - 1 \right] \right\} R^2 dR \right)}{\exp\left( -\frac{2\pi\rho}{Nk} \int_0^\infty \left\{ -\beta e^{-\beta U^{\mathrm{HS}}(R)} U^{\mathrm{HS}}(R) - \left[ e^{-\beta U^{\mathrm{HS}}(R)} - 1 \right] \right\} R^2 dR \right)} . \tag{31}$$

Since $U^{\mathrm{HS}}(R) \to 0$ as $R \to R_{\mathrm{cutoff}}$ (long-range limit), this numerical integration is well-defined and converges to a finite value. We also note that Eq. (31) can be further improved by incorporating missing higher-order many-body interactions beyond pair correlations. For example, using the Gaussian representation of coarse-grained interactions,[94] the missing cavity function and mesoscopic repulsions are expected to enhance the accuracy of Eq. (31). However, based on several bottom-up CG methods used to extract the temperature dependence of CG PMFs,[28, 139, 140] the pairwise approximation is expected to remain sufficient for quantitatively estimating the correction factor, particularly for homogeneous liquids with a one-CG-per-molecule mapping. We will validate this approximation in Sec. IV. Lastly, we note that a particular form of Eq. (31) is in fact related to the excess free energy quantity, where one can envisage its similarity to the hard sphere correction scheme suggested by Mon based on configuration integrals.[141] It would be of potential interest to establish a concrete link between these two approaches as future work.

Figure 3 summarizes the non-hard sphere CG dynamics theory that is introduced in this section. The presented non-hard sphere CG dynamics theory is composed of three separate steps: (1) Shift the CG interaction, (2) determine the hard sphere diameter from the shifted interaction and the corresponding $D_0^{\mathrm{HS}}$, and (3) calculate the hard sphere correction factor based on the pairwise interactions. By combining the correction factor to the (adjusted) hard sphere diffusion coefficient, the final CG diffusion coefficient is obtained. In the next section, we will discuss to what extent the hard sphere description breaks down and how to quantitatively account for this deviation.

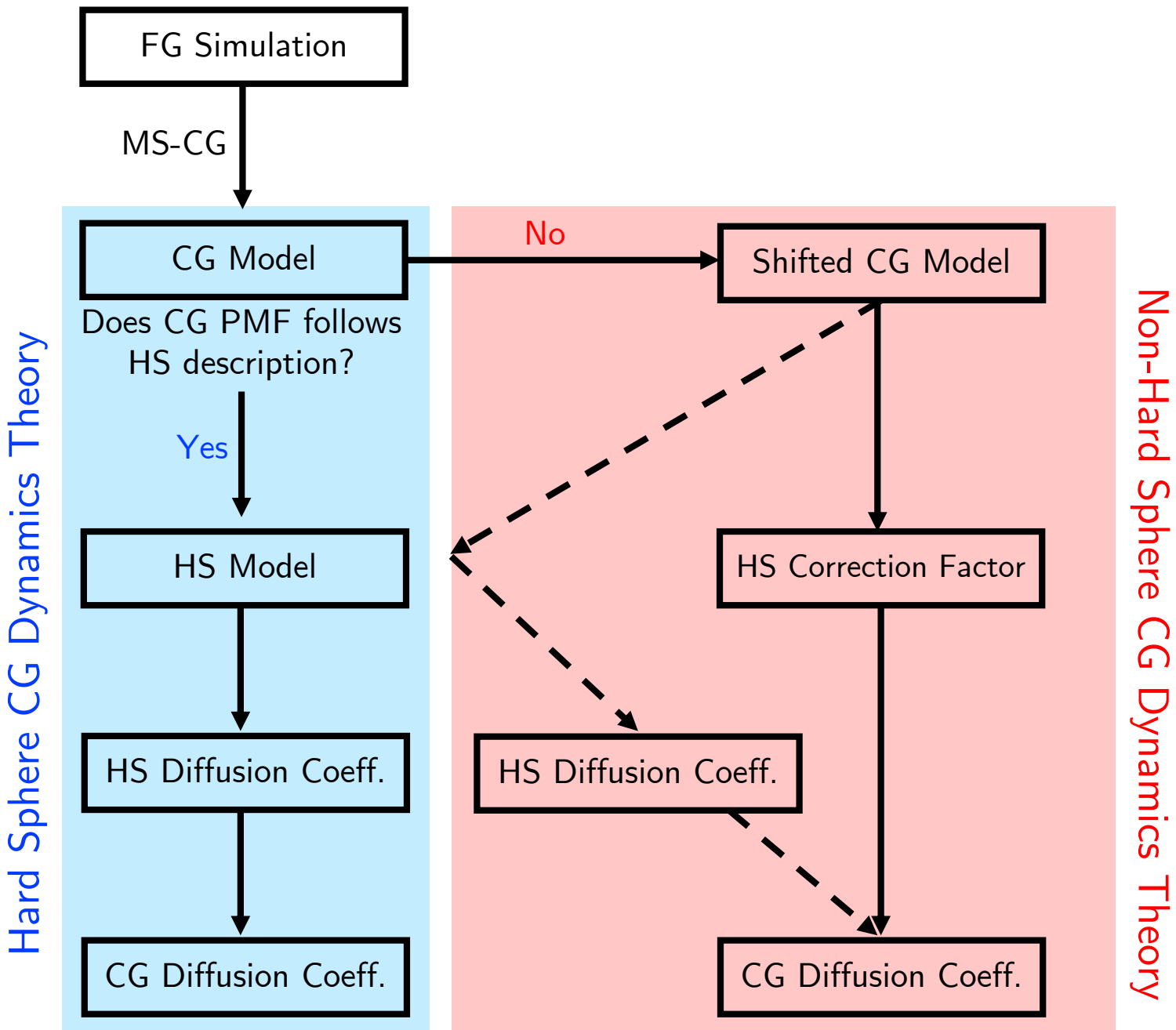

**Figure 3.** Flowchart describing the hard sphere approach used to reproduce the CG diffusion coefficient. The left block (blue) shows the hard sphere CG dynamics theory that has been proposed in the previous papers of the series. For CG systems with non-hard sphere characteristics, we propose the non-hard sphere CG dynamics theory that is depicted in the right block (red). Unlike the hard sphere theory, the degree of non-hard sphere characteristics of the CG system is corrected by the hard sphere correction factor.

### E. Computational Details

Both FG and CG simulations were carried out using the Large-scale Atomic/Molecular Massively Parallel Simulator (LAMMPS) MD engine.[142-144] All FG systems were initially prepared as 1000 molecules in the cubic box. Following the CG liquid models developed in Ref. 145, we employed the OPLS/AA[146] and the recent OPLS/CM1A[147] force fields to generate the FG trajectories. LigParGen was used to generate the input script for the LAMMPS,[148] and initial topologies of each system were randomized using the Packmol package.[149]

We first applied energy minimization to remove the unphysical instabilities from the initial structures. Then, the optimized system was heated up to the target temperature for 0.1 ns under constant *NVT* dynamics using the Nosé-Hoover thermostat.[150, 151] Target temperatures were chosen between the melting and boiling points of the system of interest. For methanol, we used a temperature range of 250–400K. A relatively narrower temperature range was used for both acetonitrile and acetone from 240 to 360K. At the target temperature, the equilibrium volume of the system was determined after constant *NPT* runs for 1 ns using the Andersen barostat.[152] Finally, the FG snapshots used to analyze the structural correlations and to parametrize the CG interactions were collected under constant *NVT* dynamics for 5 ns using the Nosé-Hoover thermostat.[150, 151] We used a coupling constant of 0.1 ps for constant *NVT* runs and 1.0 ps for the *NPT* runs, and the long-range electrostatics was computed using the Ewald summation method with a convergence tolerance of 0.0001.

From the mapped all-atom trajectories, the effective CG interactions were then obtained using the MS-CG force-matching method implemented in the OpenMSCG software.[153] We used the sixth-order B-spline functions with a resolution of 0.20Å to obtain the spline coefficients. Finally, the CG simulations were performed at the same target temperatures under constant *NVT* dynamics for 5 ns using the same thermostat settings.

## IV. Results

### A. Non-Hard Sphere CG Interactions

Figure 4 depicts the effective pairwise CG interactions obtained from the MS-CG force-matching for three different liquids. We used a temperature range from 250 to 400K at 25K intervals for water, 240 to 360K at 20K intervals for acetonitrile, and 240 to 300K at 20K intervals for acetone. For all three molecules, we observe that the CG interactions become more repulsive [positive $U(R)$ value] as the temperature increases. This is consistent with the free energy perspective of the CG interactions. Since the bottom-up CG models aim to approximate the many-body CG PMF, a free energy quantity, the CG interactions at constant volume can be thought of as the Helmholtz free energy. By applying the energy-entropy decomposition, we arrive at[95, 96, 154]

$$U_{\text{CG}}(R) = \Delta E - T\Delta S(R). \quad (32)$$

It has been shown that the pair entropic contribution to the CG PMF, $\Delta S(R)$, has a negative value over the radial domain due to the missing degrees of freedom arising from the CG process. Therefore, $U_{\text{CG}}(R)$ is expected to be more repulsive as temperature increases.

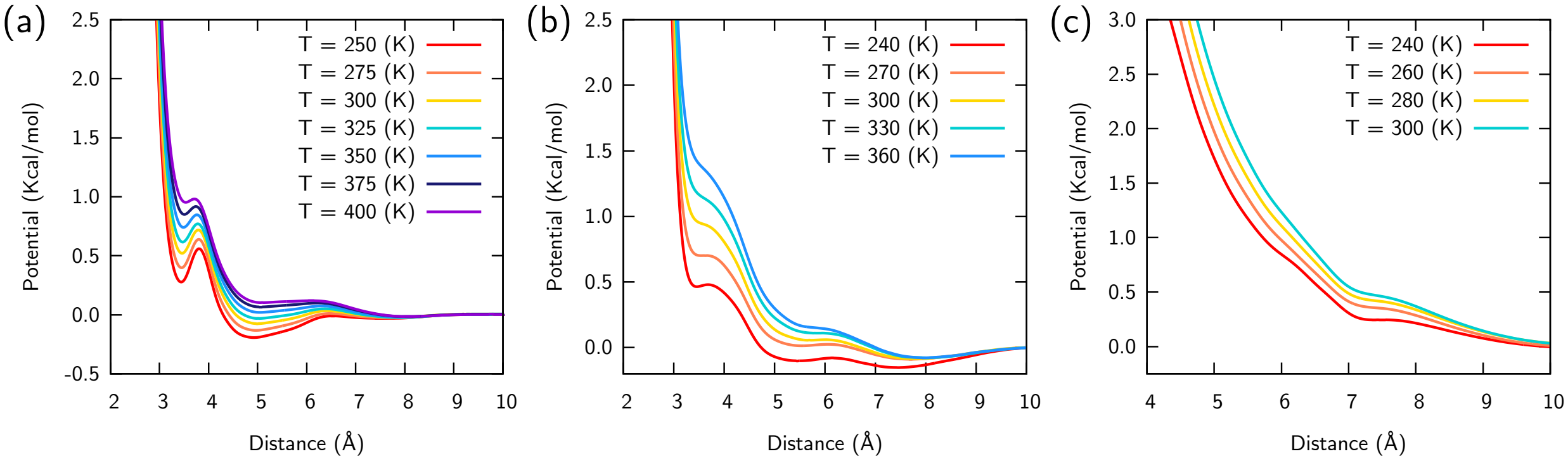

**Figure 4.** Temperature-dependent effective CG interactions of the (a) methanol, (b) acetonitrile, (c) acetone systems for various temperatures under constant *NVT* conditions. (a) Methanol-methanol CG interactions from 250 to 400K (red to purple). (b) Acetonitrile-acetonitrile CG interactions from 240 to 360K (red to blue). (c) Acetone-acetone CG interactions from 240 to 300K (red to green).

Another noticeable feature is the non-hard sphere nature of the CG interaction shown in all three liquids. In Fig. 2, we provided the reasoning for why methanol cannot be described as a hard sphere: The strong repulsive interaction at short distances does not decay to zero, and the conventional perturbation theory is not able to capture this inner-core region. For example, the CG interaction at 300K reaches zero potential energy at 4.54Å for methanol, 6.70Å for acetonitrile, and 11.10Å for acetone. However, based on the argument from Weeks-Chandler-Andersen, the zero-force region is located at 3.46Å for methanol, 3.58Å for acetonitrile, and 7.4Å for acetone. We will later naïvely calculate the Barker-Henderson diameter and see how this non-hard sphere nature could exacerbate the $D_0$ value with the hard sphere description.

Interestingly, we note that the repulsive interaction of acetone is not as steep as methanol or acetonitrile, since the interaction slowly decays up to nearly 7Å. This feature is attributed to the topology of acetone, which has a more symmetric and sphere-like structure compared to the more linear shape of methanol or acetonitrile. By mapping one molecule to its center-of-mass, we expect less repulsive pair interactions for acetone at the FG resolution. These differences in the molecular origins of the inner-core repulsion should be also considered when we map each molecule as a hard sphere entity, and we will revisit these differences in the molecular nature later in Sec. IV E.

### B. CG Results: Structure and Dynamics

Before analyzing the dynamics of CG models, we verified the fidelity of the CG models by calculating the structural correlations at different temperatures. As plotted in Fig. 5, it is apparent that the chosen temperature ranges for each molecule fall into the liquid phase. In turn, we confirmed that the optimized CG interactions are capable of recapitulating changes in pair correlations for all three liquids.

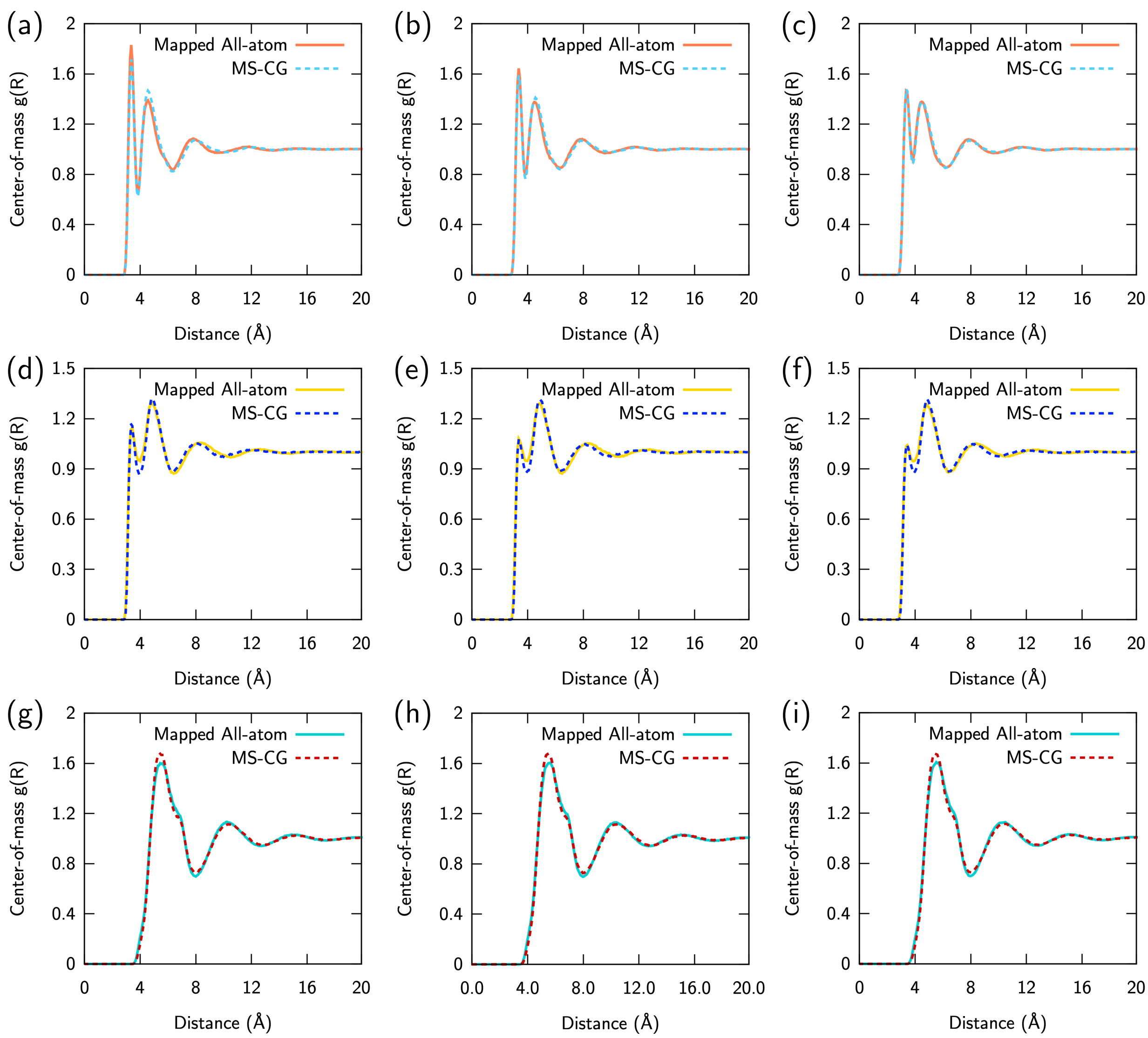

**Figure 5.** Intermolecular RDF $g(R)$ for methanol, acetonitrile, and acetone at the selected temperatures. The RDFs of methanol from the mapped atomistic (red line) and CG model (cyan dashed) are calculated at (a) 250K, (b) 300K, (c) 350K. Also, the RDFs of acetonitrile from the mapped atomistic (yellow line) and CG model (blue dashed) are calculated at (d) 270K, (e) 300K, (f) 330K. Finally, the RDFs of acetonitrile from the mapped atomistic (cyan line) and CG model (red dashed) are calculated at (g) 260K, (h) 280K, (i) 300K.

Then, we computed the self-diffusion coefficients using the mean square displacement, defined as $\langle R^2(t)\rangle \coloneqq \langle |\vec{R}(t) - \vec{R}_0|^2 \rangle = \sum_I^{N_{CG}} |\vec{R}_I(t) - \vec{R}_I(0)|^2 / N_{CG}$, via Einstein's relation

$$D = \lim_{t\to\infty} \frac{1}{6t} \langle R^2(t) \rangle \coloneqq \lim_{t\to\infty} \frac{1}{6t} \frac{1}{N_{CG}} \sum_I^{N_{CG}} |\vec{R}_I(t) - \vec{R}_I(0)|^2 . \quad (33)$$

Figure 6 lists the molecular self-diffusion coefficients from the center-of-mass configurations of methanol, acetonitrile, and acetone. The FG trajectories were manually mapped to the center-of-mass coordinates, and the diffusion coefficients were readily computed.

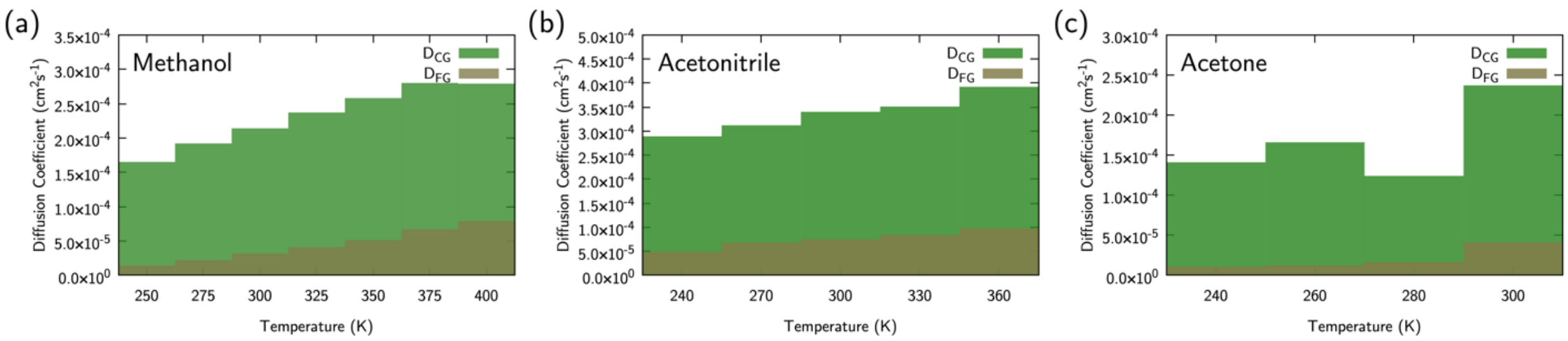

**Figure 6.** Molecular self-diffusion coefficients from the FG simulation, $D_{\text{FG}}$, and the CG simulation, $D_{\text{CG}}$, for methanol, acetonitrile, acetone at different temperature ranges: (a) methanol from 250 to 400K, (b) acetonitrile from 240 to 360K, and (c) acetone from 240 to 300K.

Two general trends observed in Fig. 6 match our expectations: (1) the diffusion coefficient increases as temperature increases, and (2) the CG dynamics are always faster than the FG dynamics. This acceleration factor is non-uniform and becomes larger as temperature increases, which can be explained by several models available in the literature, including the Speedy-Angell,[155] Vogel-Fulcher-Tammann,[156-158] Bässler,[159] and Krausser-Samwer-Zaccone laws.[160]

## C. Excess Entropy Scaling of CG Dynamics

Our particular interest in this paper is to recapitulate the $D_{\text{CG}}$ values of the non-hard sphere systems. Hence, to verify the excess entropy scaling relationship for the aforementioned CG systems, we computed the excess entropy of the CG systems based on Eq. (7) and the reduced diffusion coefficient from Fig. 6. We already provided an excess entropy scaling relationship for methanol in Paper I,[61] but we would like to examine if the scaling relationship still holds for other systems as well. For each system, the macroscopically rescaled diffusion coefficient is readily obtained as

$$D^*(\text{MeOH}) = 1.5029 \times 10^4 \times \frac{D}{\sqrt{T}}, \quad (34a)$$

$$D^*(\text{MeCN}) = 1.5312 \times 10^4 \times \frac{D}{\sqrt{T}}, \quad (34b)$$

$$D^*(\text{AcO}) = 1.5585 \times 10^4 \times \frac{D}{\sqrt{T}}, \quad (34c)$$

using the equilibrated cubic box lengths of $L_{\text{MeOH}} = 41.30\,\text{Å}$, $L_{\text{MeCN}} = 45.88\,\text{Å}$, and $L_{\text{AcO}} = 53.62\text{Å}$. Similarly, the ideal gas translational entropy for each system is given as

$$s_{\text{trn}}^{(\text{id})}(\text{MeOH}) = -\ln\left[\frac{(6.6262\times10^{-34}\text{m}^2\cdot\text{kg}\cdot\text{s}^{-1})^2}{2\pi\left(\frac{32.0\times10^{-3}\text{kg}}{6.02214\times10^{23}}\right)\times1.381\times10^{-23}\text{J}\cdot\text{K}^{-1}\times T}\right]^{\frac{3}{2}} - \ln\left[\frac{1000}{(41.30\times10^{-10}\text{m})^3}\right] + \frac{5}{2} = 3.3764 + \frac{3}{2}\ln T\,, \quad (35a)$$

$$s_{\text{trn}}^{(\text{id})}(\text{MeCN}) = 4.0638 + \frac{3}{2}\ln T\,, \quad (35b)$$

$$s_{\text{trn}}^{(\text{id})}(\text{AcO}) = 5.0516 + \frac{3}{2}\ln T\,. \quad (35c)$$

Figure 7 delineates $s_{ex}^{\text{CG}}$ over $\ln D_{\text{CG}}^{*}$ for different systems at temperatures ranging from 250 to 400K. Surprisingly, we report that the linear scaling relationship holds for different CG liquids

$$\ln D_{\text{CG}}^{*}\,(\text{MeOH}) = 0.650\times s_{ex}^{\text{FG}} - 0.861, \quad (36a)$$

$$\ln D_{\text{CG}}^{*}(\text{MeCN}) = 0.988\times s_{ex}^{\text{FG}} - 0.601, \quad (36b)$$

$$\ln D_{\text{CG}}^{*}(\text{AcO}) = 0.119\times s_{ex}^{\text{FG}} - 1.769. \quad (36c)$$

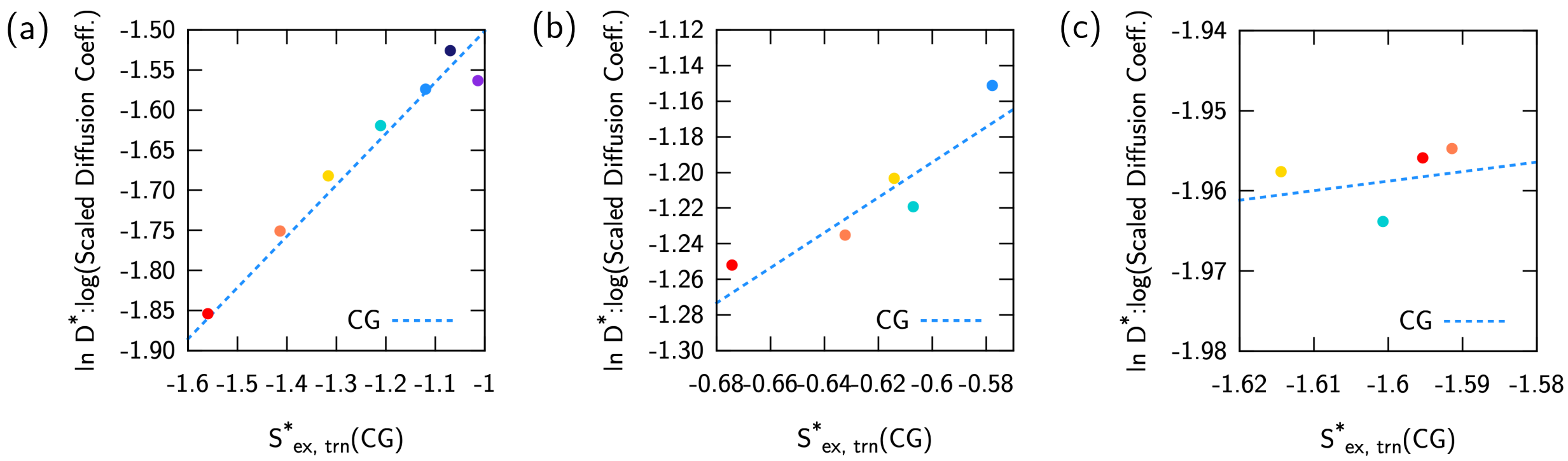

**Figure 7.** Examination of the Rosenfeld scaling for the CG liquids. (a) CG scaling relationship for methanol at different temperatures from red (250K) to purple (400K) according to Fig. 6(a), as shown in Eq. (36a). (b) CG scaling relationship for acetonitrile at different temperatures from red (240K) to blue (360K) according to Fig. 6(b), as shown in Eq. (36b). (c) CG scaling relationship for acetone at different temperatures from red (240K) to green (300K) according to Fig. 6(c), as shown in Eq. (36c). Note that for all three liquids at the different temperatures, a linear scaling relationship is observed.

As discussed in Papers I and III,[61, 63] the exponent $\alpha$ in Eq. (1) depends on the molecular nature of the system. Even though we discovered that this exponent is invariant between the FG and CG systems that represent the same molecular system, it is not transferable across different molecular systems. This explains the wide ranges of $\alpha$ values in Eq. (36a)–(36c). Instead of $\alpha$, our primary interest is the entropy-free diffusion coefficients $D_0$, which can be understood from rigorous statistical mechanics.

### D. Naïvely Estimated Hard Sphere Nature from the Barker-Henderson Approach

From Eqs. (36a)–(36c), the obtained $D_0$ suggests that the methanol, acetonitrile, and acetone systems are not able to be purely described by the hard sphere process. As a comparison, different $D_0$ values are shown in Table 2. For the Percus-Yevick EOS,[112] $D_{0,\mathrm{min}}^{\mathrm{PY}}$ at $\eta_{\mathrm{min}}$ is 0.655, but the actual $D_0$ values for three liquids are quite smaller than that. This confirms our initial observations based on the non-hard spherical CG interactions shown in Fig. 4.

To substantiate this anomaly, we naïvely computed the Barker-Henderson diameter from the CG interactions (see Table 2). For clarity, hereafter, all analysis of the CG interactions reported in this paper will be calculated at 300K. By employing Eq. (8), the naïve Barker-Henderson diameter $\sigma_{\mathrm{BH}}^{\mathrm{naïve}}$ is computed to be 3.828Å for methanol, 4.460Å for acetonitrile, and 7.680Å for acetone. It should be also noted that the $\sigma_{\mathrm{HS}}^{\mathrm{naïve}}$ value is always slightly smaller than the distance at the first zero potential energy because the contribution from $[1-\exp\left(-\beta U(R)\right)]$ is smaller than one.

From $\sigma_{\mathrm{BH}}^{\mathrm{naïve}}$, the naïve $D_0$ value was estimated under the assumption that the systems are hard spheres. The only exception among the three liquids is acetone, which consists of a relatively flat repulsion at the inner-core region. This slowly decaying repulsion results in a non-physical packing fraction $\eta$ of acetone estimated by $\sigma_{\mathrm{BH}}^{\mathrm{naïve}}$

$$\eta_{\mathrm{BH}}^{\mathrm{naïve}} = \frac{\pi}{6}{\sigma_{\mathrm{BH}}^{\mathrm{naïve}}}^3 \cdot \left(\frac{N}{V}\right) = 1000\frac{\pi}{6}\cdot\left(\frac{7.680\text{Å}}{53.62\text{Å}}\right)^3 = 1.539. \tag{37}$$

Equation (37) confirms that a repulsive nature of acetone at short distances is not purely due to the hard sphere effect, which distinguishes itself from two other liquids. Hence, we group the first type of liquids where the hard sphere effect is well encoded in the inner-core region as *hard liquids.* On the other hand, we call liquids such as acetone where a relatively soft repulsive interaction is dominant as *soft liquids.* While a naïve estimation of the hard sphere diameter or packing fraction for hard liquids is possible by employing the Barker-Henderson theory, the same approach cannot be used for soft liquids since $\eta_{\mathrm{BH}}^{\mathrm{naïve}}$ for acetone exceeds $\eta_f = 0.494$,[161] indicating that the hard sphere theory is no longer valid in this regime. In this case, certain treatments other than the Barker-Henderson approach are required to correctly represent the hard sphere characters of soft liquids; this will be discussed in the next subsection. Even though in this work we group the target systems into hard and soft liquids, similar definition should be applied for complex systems other than liquids by examining their pair interaction profiles.

For hard liquids, naïvely computed $D_0$ values from Eq. (16a) are nearly 4–8 times larger than the reference $D_0$ value, as listed in Table 2. Particularly, this difference is most pronounced in acetonitrile where the naïve treatment gives 8.6 times larger diffusion coefficients. This trend is invariant using a different EOS. As an example, Table A1 shows similar results from the Carnahan-Starling EOS.

**Table 2.** Naïvely estimated hard sphere diameter $\sigma_{\mathrm{BH}}^{\mathrm{naïve}}$ from the CG PMFs and the corresponding entropy-free diffusion coefficients $D_0$ ($\sigma_{\mathrm{BH}}^{\mathrm{naïve}}$) for soft and hard liquids. Note that the estimated $D_0$ ($\sigma_{\mathrm{BH}}^{\mathrm{naïve}}$) is larger than the minimum $D_0$ value derived from the Enskog kinetic theory using the Percus-Yevick EOS $D_{0,\mathrm{min}}^{\mathrm{PY}}(\eta_{\mathrm{min}})$. Since the naïve packing fraction for acetone is way beyond the physical limit, we leave out the estimated $D_0$ value for acetone as blank.

| System | Hard liquid | | Soft liquid |
|---|---|---|---|
| | **Methanol** | **Acetonitrile** | **Acetone** |
| $\sigma_{\mathrm{BH}}^{\mathrm{naïve}}$ | 3.828Å | 4.460Å | 7.680Å |
| $D_0$ ($\sigma_{\mathrm{BH}}^{\mathrm{naïve}}$) | 1.845 | 4.299 | - |
| *Reference* $D_0$ | 0.3866 | 0.3941 | 0.1705 |
| $D_{0,\mathrm{min}}^{\mathrm{PY}}(\eta_{\mathrm{min}})$ | | 0.655 | |

**E. Non-Hard Sphere Correction**

To effectively transform a non-hard sphere interaction into a hard sphere-like interaction, we employed Eq. (20) for the CG interactions depicted in Fig. 4. From the optimized CG interactions, $R_{\mathrm{min}}$ was chosen to satisfy $[dU/dR]|_{R=R_{\mathrm{min}}} = 0$, giving 3.46Å for methanol, 3.58Å for acetonitrile, and 7.40Å for acetone. The scaled CG interactions $U^{\mathrm{HS}}(R)$ are plotted in Fig. 8. Notably, a similar adjustment scheme for methanol was recently reported in Ref. 124 using the hybrid Barker-Henderson approximation,[122, 123] but, in that work, the corrected hard sphere diameter was used solely to estimate thermodynamic entropy via analytic equations of state without establishing a link to dynamical properties, which is a key focus of this study.

For hard liquids, the scaled CG interactions, shown in Fig. 8, distinctly represent the hard sphere character at the short-range regions. The corresponding Barker-Henderson diameter for the non-hard sphere systems $\sigma_{\mathrm{non-HS}}$ using Eq. (21) are obtained as 3.257Å for methanol, 3.175Å for acetonitrile. As expected, these values are relatively smaller than the naïve diameter $\sigma_{\mathrm{BH}}^{\mathrm{naïve}}$ by about 17.5–40.5%. Nevertheless, using this adjusted diameter, the $D_0$ values estimated from Eq. (16a) are relatively larger than the reference data (see the second row in Table 2 for the Percus-Yevick EOS and Table B1 for the Carnahan-Starling EOS), since the hard sphere correction factor is still missing.

For soft liquids, we cannot simply rely on the corrected Barker-Henderson diameter to obtain the $D_0$ values because the soft nature of inner-core repulsions is neglected. This argument is confirmed by the fact that $\sigma_{\mathrm{non-HS}} = 6.313$Å for acetone from Fig. 8 gives a packing fraction of 0.8547, which is still larger than the volume fraction at freezing. To effectively account for this soft sphere nature, we utilize the fluctuation matching approach since it is not directly dependent on the CG interactions itself. Instead, it aims to capture the overall average density fluctuations throughout the system that is guaranteed to reflect the less structured local ordering at short distances. Furthermore, a major advantage of fluctuation matching is that it effectively extracts a dynamically consistent hard sphere feature of the system. Therefore, an additional adjustment that we considered for hard liquids is not needed, and this argument is further supported by a quasi-elastic neutron scattering point of view.

From experimental efforts to understand liquid dynamics, a conventional approach based on the hard sphere description is manifested by the De Gennes narrowing relationship,[162] where the second frequency moment of the coherent dynamical structure factor $S(Q,\omega)$ with $Q \coloneqq k_i - k_f$ and $\omega = \hbar^{-1}(E_i - E_f)$ obeys the following relationship

$$\langle \omega_Q^2 \rangle = \frac{k_B T}{mS(Q)} Q^2, \tag{38}$$

where $S(Q)$ is the static structure factor. As this narrowing originates from the strong correlations between fluid particles in the nearest neighbors, Cohen *et al.* derived an alternative use of the Enskog kinetic theory to derive the half width at half maximum of the spectral width $\Delta\omega$ using the Enskog diffusion coefficient $D_E$, as given by[163-167]

$$\Delta\omega \approx \frac{D_E Q^2}{S(Q)} \cdot \frac{1}{1 - j_0(Q\sigma_{\mathrm{HS}}) + 2j_2(Q\sigma_{\mathrm{HS}})} \tag{39}$$

where the Bessel spherical functions with order 0 and 2 are denoted as $j_0$ and $j_2$, respectively. Namely, Eq. (39) can be interpreted as a fitting of the fluctuations of the dynamical structure factor using the hard sphere description. The so-called *revisited Enskog theory for the De Gennes narrowing* shares similar physics to fluctuation matching. It is further shown that Eq. (39) can be applied for the non-hard sphere system, liquid metals, in spite of strong structural deviations from a simple hard sphere result.[168-170] Hence, with this applicability of the hard sphere kinetic theory to non-hard sphere systems in mind,[171] we expect that the same analogy will take place in fluctuation matching as well. Even though the main target of fluctuation matching and the revisited Enskog theory is to obtain the structure factor and be able to extract the physical packing fraction regardless of the degree of hard sphere characteristics ("hard sphereness"), our work takes an additional step by incorporating the changes in the entropy term [Eq. (30)] within the excess entropy scaling formalism. One interesting direction extending from this analogy for future work is to explicitly determine the correspondence between these two theories.

Following the procedures described in Paper II,[62] the dimensionless compressibility $S(k=0)$ is obtained as 0.4812. Using the Percus-Yevick EOS, the fluctuation matching equation becomes

$$S(k=0)_{\mathrm{HS}}^{\mathrm{PY}} = \rho k_B T \cdot \left(-\frac{1}{V} \cdot \frac{\partial V}{\partial P}\right) = \frac{(1-\eta)^4}{(1+2\eta)^2}, \tag{40}$$

which is reduced to

$$\eta_{\mathrm{PY}} = \sqrt{S(k\to 0)_{\mathrm{HS}}^{\mathrm{PY}}} - \sqrt{S(k\to 0)_{\mathrm{HS}}^{\mathrm{PY}} + 3\sqrt{S(k\to 0)_{\mathrm{HS}}^{\mathrm{PY}}}} + 1 = 0.09300. \tag{41}$$

We note that the Carnahan-Starling EOS gives a comparable value of 0.09328 (see Appendix B). The similarity between the Barker-Henderson approach and fluctuation matching was demonstrated in Paper II for water,[62] and we further ensured that both methods provided similar packing densities for *hard liquids* by repeating the same procedures for methanol and acetonitrile as well. For hard liquids, we found that both methods agree within a 10% error for $D_0$ using $S(k=0)_{\mathrm{CG}}^{\mathrm{MeOH}} = 0.1828$ and $S(k=0)_{\mathrm{CG}}^{\mathrm{MeOH}} = 0.1489$.

In turn, both Barker-Henderson and fluctuation matching can be applied to hard liquids, but soft liquids should be only determined using fluctuation matching. Using this packing fraction, we arrive at $D_0\,(\eta_{\mathrm{non-HS}}) = 0.7964$ that accounts only for the hard sphere character of acetone. Thus, in order to recapitulate $D_0$ from the scaling relationship at the CG level, the deviation from the non-hard sphere nature should be considered as well.

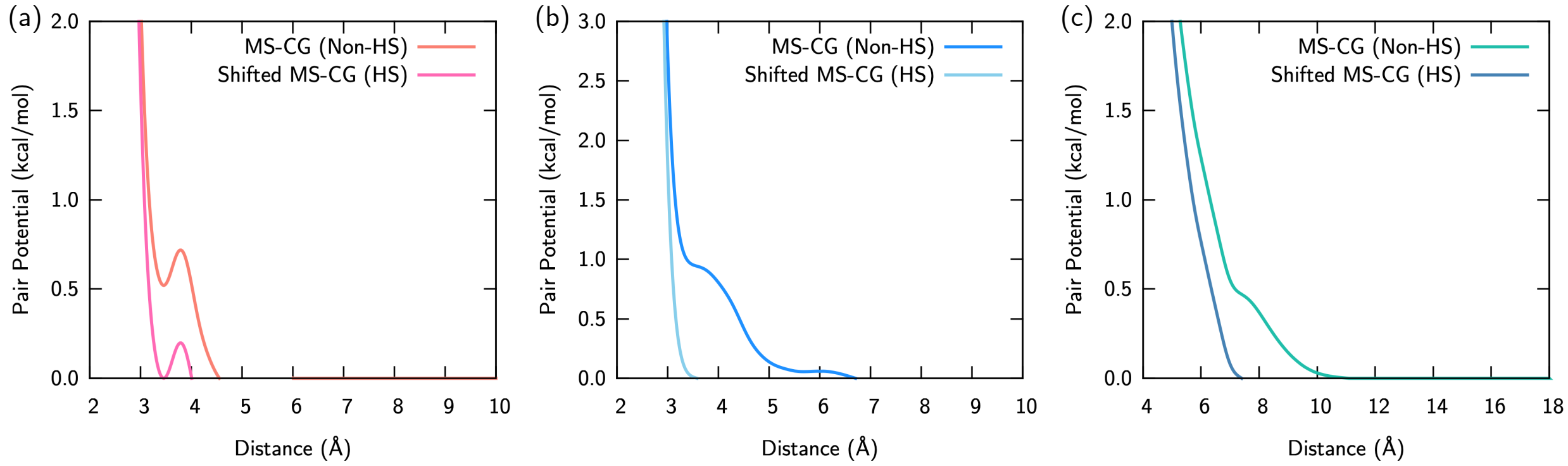

**Figure 8.** Rescaled non-hard sphere CG interactions of (a) methanol, (b) acetonitrile, and (c) acetone at 300K. The optimized CG interactions from the force-matching was adjusted to hard sphere-like interactions by rescaling $U^{\mathrm{HS}}(R) = U(R) - U(R_{\mathrm{min}})$ to partition the inner-core repulsive area defined by the sign of forces. The non-hard sphere interactions are depicted as red (methanol), blue (acetonitrile), and green (acetone), whereas the rescaled hard sphere interactions are shown as pink (methanol), sky-blue (acetonitrile), and teal (acetone), respectively.

We now report the hard sphere correction factor based on the scaled CG interactions. For methanol, using the pairwise approximation, $\exp\left(s_{ex}^{\mathrm{unshifted}}\right)$ is 0.3857, and $\exp\left(s_{ex}^{\mathrm{shifted}}\right)$ is 0.7613, giving $f$ as 0.5067. The $f < 1$ suggests that the hard sphere correction scheme is able to account for the substantially large $D_0$ from the naïve assessment. Similarly, $f$ for acetonitrile was 0.5156 using $\exp\left(s_{ex}^{\mathrm{unshifted}}\right) = 0.4170$ and $\exp\left(s_{ex}^{\mathrm{shifted}}\right) = 0.8093$. Unlike hard liquids, it is relatively less straightforward to determine the non-hard sphere effect of soft liquids. Since fluctuation matching yields an effective packing fraction that is dynamically-consistent with the averaged ensemble fluctuation, it is not directly related to the pair CG interactions. Yet, in this work, on the basis of the Weeks-Chandler-Andersen theory,[78-80] we assume that the correction factor based on the adjusting pair interaction is still valid for acetone. Nevertheless, it should be noted that much of the relationships between the non-hard sphere nature and isothermal compressibility remain unexplored, indicating that the proposed approach can be improved as a future study. With this in mind, $f$ for acetone was approximately 0.2632 from $\exp\left(s_{ex}^{\mathrm{unshifted}}\right) = 0.0295$ and $\exp\left(s_{ex}^{\mathrm{shifted}}\right) = 0.1123$. The adjusted $D_0$ values for the non-hard sphere systems are listed in Table 3 by calculating

$$D_0 = \frac{6^{\frac{1}{3}}}{4}\pi^{\frac{1}{6}}\eta^{\frac{1}{3}}\frac{(1-\eta)^2}{\eta(\eta^2-2\eta+4)}\exp\left[\frac{3(2\eta-\eta^2)}{2(1-\eta)^2}\right] \cdot \frac{\exp\left(-\frac{2\pi\rho}{Nk}\int_0^\infty\left\{-\beta e^{-\beta U(R)}U(R)-\left[e^{-\beta U(R)}-1\right]\right\}R^2\cdot dR\right)}{\exp\left(-\frac{2\pi\rho}{Nk}\int_0^\infty\left\{-\beta e^{-\beta U^{\mathrm{HS}}(R)}U^{\mathrm{HS}}(R)-\left[e^{-\beta U^{\mathrm{HS}}(R)}-1\right]\right\}R^2\cdot dR\right)} \tag{42}$$

We emphasize that the corrected diffusion coefficients provide much more accurate values in contrast to blindly assuming the hard sphere character for these systems. For methanol, the relative error was decreased by about 13-fold, whereas 4.8-fold enhancement was seen from acetonitrile. The most striking enhancement was seen from acetone where the corrected $D_0$ value gives about 16 times higher accuracy in terms of relative errors.

**Table 3.** Corrected estimated hard sphere diameter from the CG PMFs and the corresponding entropy-free diffusion coefficients $D_0$ for soft and hard liquids based on the non-hard sphere nature: the $\sigma_{\mathrm{non-HS}}$ values for soft liquids were obtained from the Barker-Henderson diameter and the $\eta_{\mathrm{non-HS}}$ values for hard liquids were obtained from fluctuation matching. Then, the resultant $D_0$ (non-HS) values are corrected by including the correction factor for changes in the entropic contributions. Here, we used the Percus-Yevick EOS for treating the hard sphere system. For the Carnahan-Starling EOS, see Table B1.

| System | *Hard liquid* | | *Soft liquid* |
|---|---|---|---|
| | **Methanol** | **Acetonitrile** | **Acetone** |
| *Effective hard sphere measure* | $\sigma_{\mathrm{non-HS}}$ | | $\eta_{\mathrm{non-HS}}$ |
| | 3.257Å | 3.175Å | 0.09300 |
| $D_0$ (non-HS) | 0.7137 | 0.6574 | 0.7964 |
| Corrected $D_0$ (non-HS) | 0.3616 | 0.3390 | 0.2096 |
| *Reference* $D_0$ | 0.3866 | 0.3941 | 0.1705 |

To summarize, the developed framework is illustrated with a flowchart in Fig. 9.

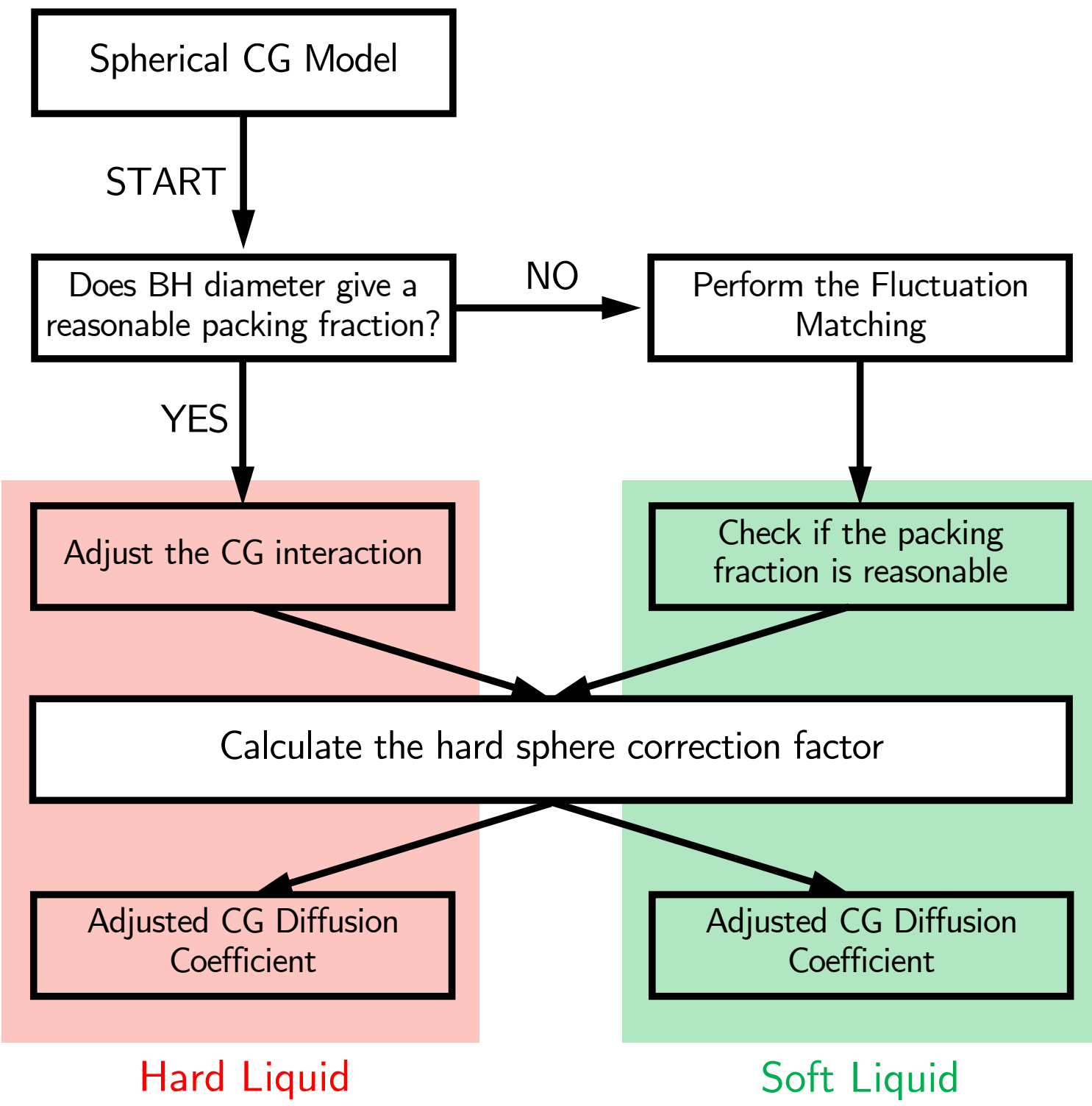

**Figure 9.** Flowchart summarizing the non-hard sphere CG dynamics theory presented in this work. For the CG systems with spherical CG interactions, we first determine if the naïvely determined BH diameter provides a physical packing fraction of hard liquids (left block). If not, we alternatively utilize the fluctuation matching introduced in Ref. 62 to determine the naïve packing fraction of soft liquids (right block). The hard sphere correction factor is then calculated based on the hardness/softness of the CG system. Using the computed correction factor, the hard sphere description of the CG diffusion is then adjusted.

### F. Extension to Generalized Non-Hard Sphere Liquids: Theoretical Foundations

In the last subsection, the hard sphere correction factors for each CG system are separately determined based on the specific CG interaction profile of the system of interest. Despite its success for a few small liquids, the proposed theory might lack applicability to other non-hard sphere CG systems. Hence, we seek to extend the current framework to a more generalized description of liquids and other relatively simple systems under the condition that the CG interactions can be reduced as an analytical interaction form. Yet, given the bottom-up nature of the CG models, we would like to emphasize that it is an extremely difficult task to determine the bottom-up CG interactions, which, in principle, cannot be determined as a single term as they are parametrized directly from the reference trajectories and vary over different system conditions.[14, 96, 172-175]

In this section, we provide an alternative approach as a continuation of our recent studies.[81, 82, 94] While maintaining the optimized CG interactions from various methods (force-matching in this case), we aim to dissect bottom-up CG interactions by decomposing them into different contributions with different physical origins. Thus, after understanding and addressing each sub-interaction correctly, the generalized CG interaction can be constructed by parametrizing each component and recombining them.

The first step to partitioning each sub-interaction can be understood from the classical perturbation theory of liquids, allowing for partitioning the hard sphere repulsion at the inner-core region. After extracting out the hard sphere repulsive interactions, the remaining perturbative interactions are non-trivial, and thus a quantitative model based on the correct physical intuition is required. One particular example to note in this similar vein is the pairwise isotropic interaction model between single-site CG water molecules, known as the continuous shouldered well (CSW) model.[176] The CSW model incorporates the hard sphere repulsive interaction with the continuous shouldered well interaction on the basis of the Stell-Hemmer core-softened potential that has been suggested to capture the phase transition behavior.[177, 178] In turn, the CSW model for water $U_{\mathrm{CSW}}^{\mathrm{H_2O}}(R)$ is defined as

$$U_{\mathrm{CSW}}^{\mathrm{H_2O}}(R) = \frac{U_R}{1+\exp\left(\frac{\Delta(R-R_R)}{a}\right)} - U_A \cdot \exp\left[-\frac{(R-R_A)^2}{2\delta_A^2}\right] + \left(\frac{a}{R}\right)^{24}. \qquad (43)$$

In Eq. (43), the $(a/R)^{24}$ term corresponds to the hard sphere repulsion for water in addition to two non-trivial interactions with interaction strength represented by $U_R$ and $U_A$. Other parameters $a, \Delta, R_R, R_A,$ and $\delta_A$ are introduced to correctly model the shape of the interaction profile. It has been shown that the CSW model for water is capable of recapitulating the structural correlations and anomalies exhibited by water at lower temperatures.[176, 179, 180] Especially, despite its *ad hoc* nature, we recently demonstrated a link between this model and the bottom-up CG model for water.[82] Therefore, a natural extension is to generalize Eq. (43) for the CSW-like model for other CG liquids with the following interaction form

$$U_{\mathrm{CSW}}(R) = \left(\frac{a}{R}\right)^{\alpha} + \frac{U_R}{1+\exp\left[\frac{R-R_R}{\delta_R}\right]} - U_A \cdot \exp\left[-\frac{(R-R_A)^2}{2\delta_A^2}\right]. \qquad (44)$$

Yet, due to the limitation of the CSW model itself, Eq. (44) can be applied to CG interactions with a continuous shouldered shape. Among the three liquids studied in this work, acetonitrile is particularly suitable for the CSW description, and we fitted Eq. (44) to the acetonitrile interactions shown in Fig. 10(a)–(b). We note that $U_A$ becomes zero in this case because there is no attractive acetonitrile interactions. By extracting the inner-core repulsion from the relatively long-range repulsion described by the $U_R/[1+\exp[(R-R_R)/\delta_R]]$ term, it is immediately apparent that the non-hard sphere nature originates from the existence of this additional repulsive interaction.

To enhance the expressiveness of the generalized CG model, an alternative choice is to modify the interaction form of the CSW model shown in Eq. (44) into the Gaussian double well (GDW) model in order to keep both repulsion and attraction

$$U_{\mathrm{GDW}}(R) = \left(\frac{a}{R}\right)^{\alpha} + U_R \cdot \exp\left[-\frac{(R-R_R)^2}{2\delta_R^2}\right] - U_A \cdot \exp\left[-\frac{(R-R_A)^2}{2\delta_A^2}\right], \qquad (45)$$

where the inner-core interaction $(a/R)^{\alpha}$ is invariant, and the repulsive and attractive Gaussian interactions are expressed as the strengths $U_R$, $U_A$ with the variances $\delta_R^2$, $\delta_A^2$, respectively. Unlike the CSW model, the GDW model is expected to capture not only the shouldered well, but also the various interaction profiles such as single or double wells. In terms of a statistical mechanical point of view as well as applicability, the approximation of effective interactions as Gaussians is reasonable since it is rooted in the classical density functional theory (DFT),[181-184] where the density-dependent interactions can be faithfully approximated using the Gaussian kernels. Relatedly, the double-Gaussian interaction models[185] as well as the Gaussian-core model[186] have been widely employed in the polymeric systems.[187-191] The main difference between the GDW model and the Gaussian-core model is that the GDW model is built upon the hard sphere nature of systems at the microscopic regime, such that a strongly repulsive hard sphere nature is better represented as a polynomial function rather than a smoothly decaying Gaussian function.

Yet, we would like to emphasize that the formulation of Eq. (45) is still quite *ad hoc*, as the GDW is neither purely derived from statistical mechanical theory nor from CSW. To summarize, we first generalize three different liquids studied in this work in terms of the CSW and GDW models and evaluate the corrected $D_0$ values using this generalized *ad hoc* description. After assessing the performance of the designed *ad hoc* CG models in Sec. IV G-H and Appendices C-D, we aim to provide more rigorous approach to formulate the generalized CG models based on our recent finding in Sec. I.

**G. Extension to Generalized Non-Hard Sphere Liquids: Gaussian Potential**

We now present our simplified Gaussian model incorporating the hard sphere nature of the liquid using Eq. (45). In practice, we fitted $a, \alpha, U_R, U_A, R_R, R_A, \delta_R$ and $\delta_A$ to the MS-CG interactions from force-matching. To ensure numerical stability for parametrization, we assumed that $b \gg 1$ (this term is 24 for the CSW model), and $a < R_R < R_A$, thus the $(a/R)^{\alpha}$ term vanishes at the Gaussian regimes. This assumption provides a numerically stable parametrization process to fit the two Gaussian functions. Following the approach outlined in Ref. 94, numerical settings for the parametrization were implemented in MATLAB R2019b.[192]

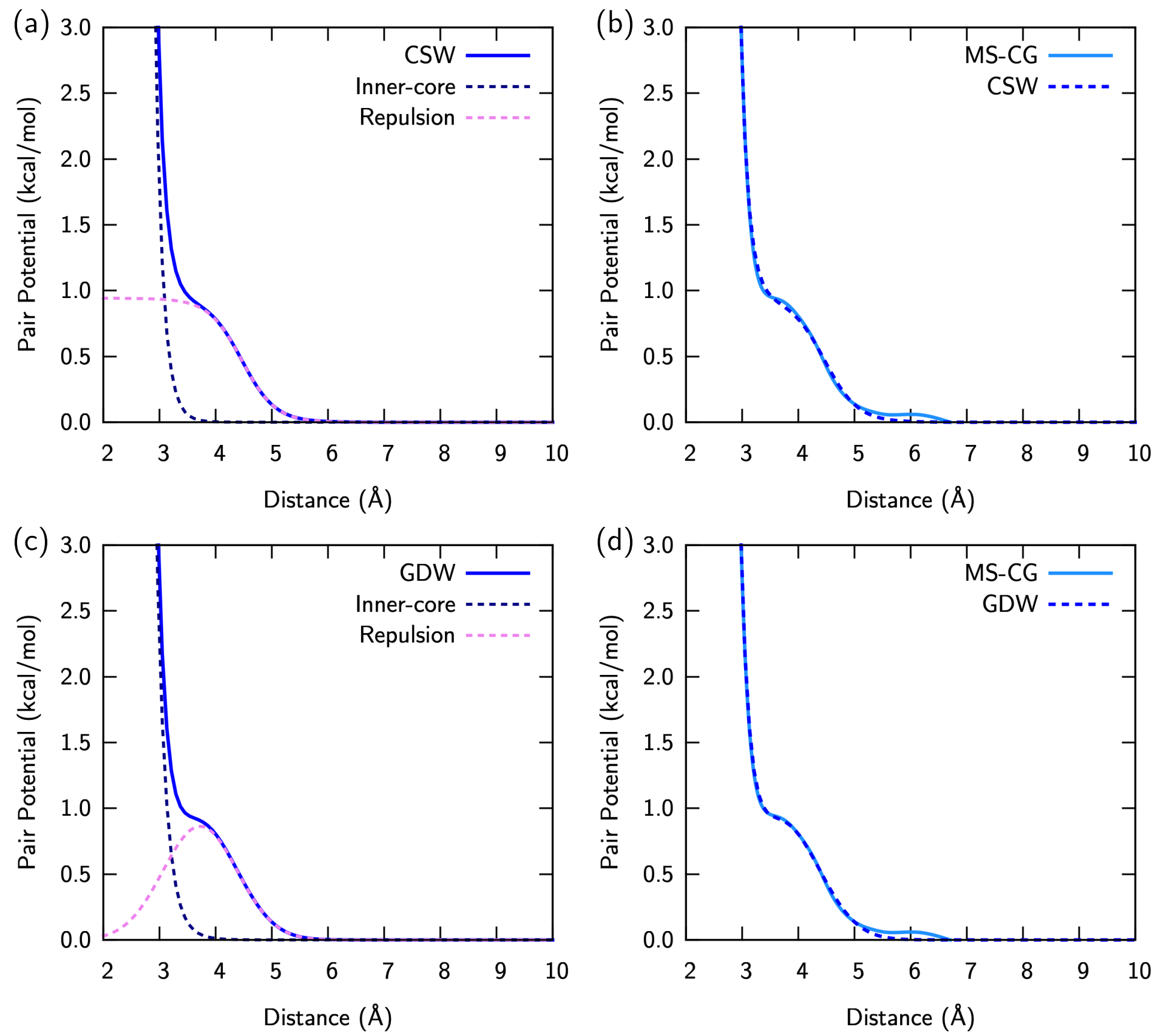

**Figure 10.** Assessing the generalized bottom-up CG model via simple parametrization: (a-b) CSW and (c-d) GDW for acetonitrile. Contributions from the inner-core repulsion (navy dashed) and additional repulsion (pink dashed) of the CSW and GDW models (blue line) are shown in (a) and (c), respectively. Overall performances in terms of reproducing the CG interactions are demonstrated in (b) for the CSW model and (d) for the GDW model (blue dashed) in comparison to the reference MS-CG interaction (sky blue line).

Figure 10 suggests that both of the parametrized CSW and GDW models are capable of reproducing the CG interaction of acetonitrile. Interestingly, the GDW description is more accurate than the CSW at the short-range regime (< 5Å). To further assess the accuracy of the GDW model, we utilize the relative entropy measure $S_{\text{rel}}$, also known as the Kullback-Leibler divergence,[193] to characterize the accuracy of the parametrized model in comparison with the force-matched reference following the analysis provided in Ref. 194.

$$S_{\text{rel}}\left(U_{\text{model}}(\mathbf{R})\right) = \int P_{\text{ref}}(\mathbf{R}) \ln \frac{P_{\text{ref}}(\mathbf{R})}{P_{\text{model}}(\mathbf{R})} d\mathbf{R}. \tag{46}$$

In Eq. (46), the probability distributions for different models are determined by their CG interactions:

$$P_{\text{ref}}(\mathbf{R}) \coloneqq \frac{\exp\left(-\beta U_{\text{ref}}(\mathbf{R})\right)}{\int d\mathbf{R} \exp\left(-\beta U_{\text{ref}}(\mathbf{R})\right)} \tag{47a}$$

$$P_{\text{model}}(\mathbf{R}) \coloneqq \frac{\exp\left(-\beta U_{\text{model}}(\mathbf{R})\right)}{\int d\mathbf{R} \exp\left(-\beta U_{\text{model}}(\mathbf{R})\right)} \tag{47b}$$

In turn, $S_{\text{rel}}\big(U_{\text{model}}(\mathbf{R})\big)$ obtained from Eqs. (46) and (47) sheds light on the information loss incurred upon different parametrizations (CSW or GDW) with the interaction parameter $U_{\text{model}}(\mathbf{R})$, allowing for evaluating the accuracy of the parametrized model.

**Table 4.** Evaluating the accuracy of the parametrized model by employing the relative entropy (or Kullback-Leibler divergence[193]) as a measure. Methanol and acetone can be only parametrized using the GDW model, as explained in the next subsection.

| System | Relative Entropy Values | | |
|---|---|---|---|
| | **CSW** ($S_{\text{rel}}^{\text{CSW}}$) | **GDW** ($S_{\text{rel}}^{\text{GDW}}$) | $S_{\text{rel}}^{\text{CSW}}/S_{\text{rel}}^{\text{GDW}}$ |
| Acetonitrile | 2.051 | 0.6699 | 3.061 |
| Methanol | - | 0.1149 | - |
| Acetone | | 0.5916 | |

Table 4 supports our observation from Fig. 10 that the GDW model is three times more accurate than the CSW model, implying that the Gaussian basis set might be a better choice not only due to its physical connection to liquids but also due to its precision in reproducing the force-matched interactions.

### H. Generalized Non-Hard Sphere Liquids: Gaussian Double Well Interaction

We now apply the parametrization scheme introduced in Sec. IV F–G to other liquids: methanol and acetone. Unlike acetonitrile, methanol no longer has a shouldered shape at the repulsive region near 3.5–4Å. Because of the existence of this well, the CDW model is not able to reproduce the methanol interaction. From the molecular perspective, this repulsion can be understood from the relative T-shape of the methanol pairs experiencing a rod-like interaction.[195] As depicted in Fig. 11, Gaussian basis sets are capable of encoding this behavior as a collective local Gaussian repulsion. Interestingly, the inner-core repulsion coefficient obtained for methanol, $\alpha = 14.78$, is consistent with the fitted inverse power law coefficients of Lennard-Jones liquids reported in Ref. 196. The validity of this form of repulsive interactions was further confirmed due to capturing structural and dynamical features,[196] including point-to-set length scales,[197] of Kob-Andersen binary Lennard-Jones liquid systems.[198] This agreement explains the observation that the hard-core nature of the GDW methanol model resembles the repulsive part of the Lennard-Jones interaction and further confirms the existence of hard-core repulsive characteristics in molecular pair interactions.

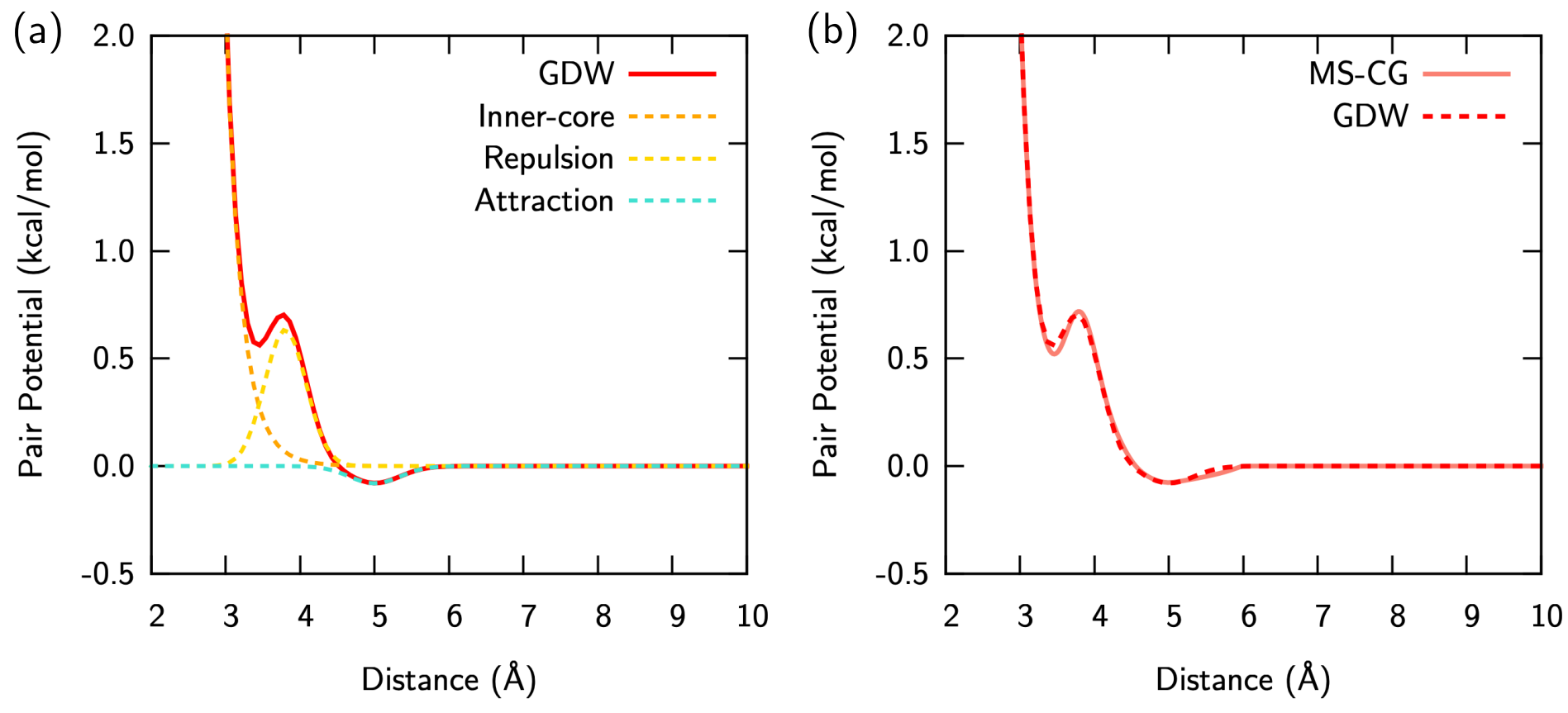

**Figure 11.** Parametrized Gaussian liquid model for methanol with double wells (GDW): (a) contributions from the inner-core repulsion (orange dashed), additional repulsion (yellow dashed), and attraction (teal dashed) in comparison to the overall GDW interaction (red line), (b) overall performance of the GDW interaction (red dashed) by comparing to the MS-CG reference (red line).

Furthermore, we tested the efficacy of the CSW and GDW models for the acetone interaction. Even though a slope of inner-core repulsions was steep for methanol ($\alpha$ =14.78) and acetonitrile ($\alpha$ =22.06), Fig. 4(c) indicates that acetone has a much slower decaying repulsion than the other two liquids. Therefore, because of the nonzero intensities of the $(a/R)^{\alpha}$ term at large distances, the parametrization of the GDW model might be problematic. Hence, we first fit the $(a/R)^{\alpha}$ term at short distances before the local minimum at 7.4Å, and then fit the rest of Gaussians to the residual. The fitted slope $\alpha = 5.49$ is even smaller than the conventional Lennard-Jones interaction, revealing the nature of soft liquids. By doing so, we find that the CSW model is not expressive enough to provide a converged solution. Yet, the GDW model can be fitted to the residuals, indicating the wide transferability of the Gaussian basis sets. This argument is further substantiated by calculating the relative entropy value. The relative entropies of the GDW model for methanol and acetone are also listed in Table 4. Note that we cannot compare the accuracy of the GDW model to the CSW model because the CSW model cannot be fitted to this case. Nevertheless, under the same GDW form, the relative entropy values for methanol and acetone are comparable to that of the acetonitrile model, implying the GDW model maintains its accuracy over different molecular systems.

**Table 5.** Optimized parameters for the simplified description of methanol, acetonitrile, and acetone. Here, we consider both CSW and GDW models, in which the model interactions are given as Eqs. (44) and (45), respectively. However, the CSW model can be only fitted to acetonitrile, whereas the GDW models can be widely applied for all three systems. Interaction strengths $U_R$ and $U_A$ are in kcal/mol. The GDW model for methanol has an additional attraction term based on the CG interaction profile.

| **Model** | | **Inner-core** | | **Repulsive** | | | **Attractive** | | |
|---|---|---|---|---|---|---|---|---|---|
| | | $a$ (Å) | $\alpha$ | $U_R$ | $R_R$ (Å) | $\delta_R$ (Å) | $U_A$ | $R_A$ (Å) | $\delta_A$ (Å) |
| Methanol | *GDW* | 3.159 | 14.78 | 0.6338 | 3.800 | 0.2796 | 0.08 | 5.00 | 0.3386 |
| Acetonitrile | *CSW* | 3.077 | 22.06 | 0.9416 | 4.467 | 0.2964 | - | | |
| | *GDW* | 3.136 | 18.29 | 0.864 | 3.725 | 0.6630 | - | | |
| Acetone | *GDW* | 6.317 | 5.49 | 0.1048 | 7.954 | 0.4139 | - | | |

The final optimized parameters for both CSW and GDW models are provided in Table 5. As can be discerned from Fig. 12, the local minimum of the CG potential, where the CG pair force becomes zero, is well described by having a relatively small Gaussian bump along with the non-zero contribution from the repulsive interaction.

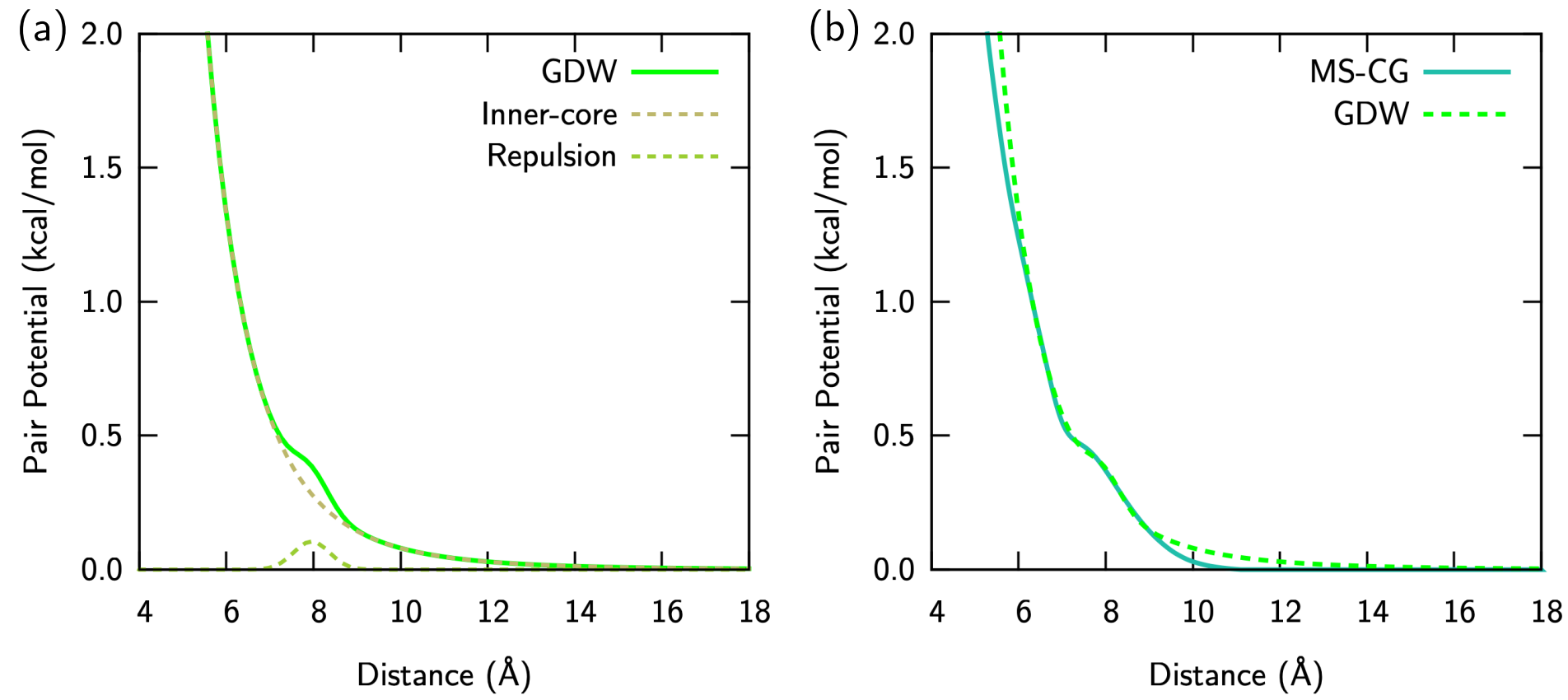

**Figure 12.** Parametrized Gaussian liquid model for acetone with double wells (GDW): (a) contributions from the inner-core repulsion (olive dashed) and additional repulsion (green dashed) in comparison to the overall GDW interaction (lime line), (b) overall performance of the GDW interaction (lime dashed) by comparing to the MS-CG reference (teal line).

For additional details about the success of the GDW model, Appendices C and D expand the applicability of the Gaussian basis sets to a broader range of CG liquids by introducing multiple Gaussian interactions as a generalized Gaussian model (gGM).

**I. Bottom-up Many-Body Gaussian Representation Models**

While the generalized *ad hoc* Gaussian models can readily approximate the MS-CG interactions and also provide a fair estimate for the $D_0$ values for non-hard sphere systems, these *ad hoc* models might lack in the rigorous physics that underlie the nature of the sub-interactions. This is because the CSW model was primarily designed to capture the shouldered well of the water interaction in an *ad hoc* manner, and the GDW model is a flexible variant of the CSW model, with both models failing to capture many-body correlations. In order to resolve this limitation, we have recently addressed this issue and provided a more rigorous connection between CG interactions and liquid physics by developing the *bottom-up many-body Gaussian representation models*.[94] Hence, in this subsection, we would like to briefly review the Gaussian representation theory and apply the developed framework in this paper to the Gaussian representation models from bottom-up.

In a nutshell, the Gaussian representation theory starts by deconvoluting the complex many-body CG interactions using both classical perturbation and density functional theoretical approaches. From the many-body expansion, the many-body CG PMF can be expressed as[199]

$$U_{\mathrm{CG}}(R) = U_2(R) + U_{\mathrm{MB}}(R), \tag{48}$$

where $U_2(R)$ is the two-body interactions, and $U_{\mathrm{MB}}(R)$ are the embedded higher-order interactions. After extracting the short-range hard sphere repulsion $U_{\mathrm{HS}}(R)$ through

$$U_2(R) = U_{\mathrm{HS}}(R) + U_2'(R), \tag{49}$$

we introduce Gaussian basis sets to decompose $U_2'(R)$ into $K$ different Gaussian interactions $G_K(\epsilon_K, \mu_K, \sigma_K)$,

$$U_2'(R) = \sum_K G_K(\epsilon_K, \mu_K, \sigma_K)\,. \tag{50}$$

Once we choose the Gaussian interactions with minimal overlap, this design results in separate Gaussian interactions that give rise to separable local densities $\rho(R)$,

$$\rho(R) = \frac{1}{V}\sum_{IJ} \delta(R - R_{IJ}) = \frac{1}{V} 4\pi\rho \int_0^R R'^2 g(R') dR' \approx \sum_{i\ (\bar{R}_{i-1}<R<\bar{R}_i)} \rho_i(\bar{R}_i, R), \tag{51}$$

where $g(R')$ is the RDF related to $U_2(R)$ via the reversible work theorem.[200] In other words, Eqs. (50) and (51) provide a physical explanation of using Gaussian interactions in the GDW model. Finally, the many-body nature presented in Eq. (48) in Appendix C can be approximated as a (single) Gaussian-core interaction by expressing the cavity correlation function[201] as

$$U_{\mathrm{MB}}(R) \coloneqq A_{\mathrm{MB}} \exp\left(-\frac{R^2}{2\delta_{MB}^2}\right), \tag{52}$$

where this smoothly decaying function is consistent with the dissipative particle dynamics at the mesoscopic description.[202-204] A detailed theoretical derivation of Eq. (52) from a cavity correlation function formalism is provided in Ref. 201. While the Gaussian sub-interactions are mostly localized interactions that originate from the local pair ordering, the slowly-decaying many-body term is a long-range repulsive interaction, which can be understood from the collective behavior of renormalized CG blobs at a lower resolution. Finally, we arrive at the interaction form for the many-body Gaussian representation model $U_{\mathrm{MB-GRM}}(R)$:

$$U_{\mathrm{MB-GRM}}(R) = \left(\frac{R_0}{R}\right)^{\delta_0} + \sum_K G_K(\epsilon_K, \mu_K, \sigma_K) + A_{\mathrm{MB}} \exp\left(-\frac{R^2}{2\delta_{MB}^2}\right). \tag{53}$$

By comparing Eq. (48) with Eq. (53), we find an alternative interpretation for the generalized Gaussian model that $U_{\mathrm{gGM}}(R)$ [Eq. (C1)] is a pairwise approximated form of $U_{\mathrm{MB-GRM}}(R)$. By introducing a pairwise approximation, the many-body density interaction $A_{\mathrm{MB}} \exp(-R^2/2\delta_{MB}^2)$ smears into the sub-Gaussians expressed in Eq. (48).

Even though the Gaussian representation models contain more parameters than the previous *ad hoc* models (e.g., CSW and GDW), these parameters can be viewed as simple EOSs that depend on their thermodynamic state points (temperature and system density). In other words, we demonstrated that the Gaussian representation models not only provide accurate CG modeling but also greatly enhances transferability over a vast range of thermodynamic state points by determining each of sub-interactions at different system conditions.

Especially, in Ref. 94, we have determined the effective Gaussian interaction form for methanol and acetonitrile, given by the following form:

$$U_{\mathrm{MeOH}}(R) = a_{\mathrm{MeOH}}(\rho, T) \exp\left(-\frac{(R-3.8)^2}{c_{\mathrm{MeOH}}^2(\rho, T)}\right) + \left(\frac{d_{\mathrm{MeOH}}(\rho, T)}{R}\right)^{15}$$
$$-f_{\mathrm{MeOH}}(T) \exp\left(-\frac{\left(R - g_{MeOH}(\rho, T)\right)^2}{h_{MeOH}^2(\rho)}\right) + k_{MeOH}(\rho, T)\, exp\left(-\frac{R^2}{15}\right), \tag{54}$$

$$U_{\mathrm{MeCN}}(R) = a_{\mathrm{MeCN}}(\rho, T) \exp\left(-\frac{(R-3.8)^2}{0.4485^2}\right) + \left(\frac{d_{\mathrm{MeCN}}(\rho, T)}{R}\right)^{15}$$
$$-f_{\mathrm{MeCN}}(\rho) \exp\left(-\frac{\left(R - g_{\mathrm{MeCN}}(\rho, T)\right)^2}{0.3759^2}\right) + k_{\mathrm{MeCN}}(\rho, T) \exp\left(-\frac{R^2}{15}\right), \tag{55}$$

where the coefficients $a_{\mathrm{MeOH}}, c_{\mathrm{MeOH}}, d_{\mathrm{MeOH}}, f_{\mathrm{MeOH}}, g_{\mathrm{MeOH}}, h_{\mathrm{MeOH}}$, and $k_{\mathrm{MeOH}}$ for methanol and to $a_{\mathrm{MeCN}}, d_{\mathrm{MeCN}}, f_{\mathrm{MeCN}}, g_{\mathrm{MeCN}}$, and $k_{\mathrm{MeCN}}$ for acetonitrile were determined by fitting the CG interaction over 111 different thermodynamic state points. Determining these EOSs for various sub-interaction terms requires extensive parametrization, and readers who are interested in the parametrization details may refer to Ref. 94. We obtained the following series of coefficients as listed in Table 6 using the system conditions for methanol and acetonitrile used in work: $\rho_{\mathrm{MeOH}} = 0.7553\mathrm{g\cdot cm^{-3}}$ and $\rho_{\mathrm{MeCN}} = 0.7058\mathrm{g\cdot cm^{-3}}$ at $T = 300\mathrm{K}$.

Additionally, as shown in Table 6, we utilized the developed parametrization protocol to determine the coefficients for the EOSs for the acetone interactions at $\rho_{\mathrm{AcO}} = 0.6256\ \mathrm{g\cdot cm^{-3}}$ to be

$$U_{\mathrm{AcO}}(R) = a_{\mathrm{AcO}}(\rho,T)\exp\left(-\frac{(R-7.941)^2}{c_{\mathrm{AcO}}^2(\rho,T)}\right) + \left(\frac{d_{\mathrm{AcO}}(\rho,T)}{R}\right)^4 \\ -f_{\mathrm{AcO}}(\rho,T)\exp\left(-\frac{\left(R-g_{\mathrm{AcO}}(\rho,T)\right)^2}{h_{\mathrm{AcO}}^2(\rho)}\right) + k_{\mathrm{AcO}}(\rho,T)\exp\left(-\frac{R^2}{15}\right), \tag{56}$$

where the coefficients in Eq. (56) are similarly defined as those in Eqs. (54) and (55) for methanol and acetonitrile, respectively.

**Table 6.** Fitted parameters for the many-body Gaussian representation of CG models for methanol [Eq. (54)], acetonitrile [Eq. (55)], and acetone [Eq. (56)]. Note that the values for methanol and acetonitrile are obtained by extrapolating the EOSs published in Ref. 94.

| **System** | **Fitted Parameters for Gaussian Representation** | | | | | | |
|---|---|---|---|---|---|---|---|
| | $a$ (kcal/mol) | $c$ (Å) | $d$ (Å) | $f$ (kcal/mol) | $g$ (Å) | $h$ (Å) | $k$ (kcal/mol) |
| Methanol | 0.5909 | 0.3507 | 3.1463 | 0.1129 | 5.1129 | 0.3166 | 0.1471 |
| Acetonitrile | 0.6843 | 0.7657 | 3.1620 | $9.715\times10^{-2}$ | 7.9140 | 1.021 | 0.3944 |
| Acetone | $7.7706\times10^{-3}$ | $4.839\times10^{-4}$ | 5.6379 | $5.458\times10^{-2}$ | 13.750 | 4.034 | 4.298 |

Appendix D demonstrates the hard sphere correction scheme starting from $U_{\mathrm{gGM}}(R)$ to obtain $U_{\mathrm{HS-gGM}}(R)$, as illustrated in Eqs. (D1)–(D5). A similar strategy can be also applied to the many-body Gaussian representation model for determining $U_{\mathrm{HS-MB-GRM}}(R)$

$$U_{\mathrm{HS-MB-GRM}}(R) = \left(\frac{R_0}{R}\right)^{\delta_0} + \sum_K G_K(\epsilon_K,\mu_K,\sigma_K) + A_{\mathrm{MB}}\exp\left(-\frac{R^2}{2\delta_{MB}^2}\right) - U_{\mathrm{MB-GRM}}(R_{\mathrm{HS}}), \tag{57}$$

where $R_{\mathrm{HS}}$ can be obtained by utilizing a similar approach as in Eq. (D4) to Eq. (57). Finally, we calculate the correction factor $f$ for the many-body Gaussian representation

$$f \approx \frac{\exp\left(-\frac{2\pi\rho}{Nk}\int_0^\infty\left\{-\beta e^{-\beta U_{\mathrm{MB-GRM}}(R)}U_{\mathrm{MB-GRM}}(R) - \left[e^{-\beta U_{\mathrm{MB-GRM}}(R)} - 1\right]\right\}R^2\cdot dR\right)}{\exp\left(-\frac{2\pi\rho}{Nk}\int_0^\infty\left\{-\beta e^{-\beta U_{\mathrm{HS-MB-GRM}}(R)}U_{\mathrm{HS-MB-GRM}}(R) - \left[e^{-\beta U_{\mathrm{HS-MB-GRM}}(R)} - 1\right]\right\}R^2\cdot dR\right)}. \tag{58}$$

Table 7 summarizes the key results of this paper. As expected, the more accurate many-body Gaussian representation models give lower error than the GDW or the generalized Gaussian models for most of the cases, as shown in Table D1 in Appendix D. In comparison to using the parametrized MS-CG interactions ("actual model" in Table D1), the many-body Gaussian representation models yield more accurate $D_0$ values within 12.40%, whereas the generalized Gaussian models are less accurate with relative errors of 29.66%. Nevertheless, we would like to point out that the relative error of these *ad hoc* models is still fairly good enough to estimate the correction factor. In such a way, combining the extended CG dynamics theory for non-hard sphere systems with the simplified CG description can provide an efficient approach for reproducing both structural and dynamical information of the CG model.

**Table 7.** Performance of reduced CG models compared to the MS-CG models for the calculation of the corrected $D_0$ values. Reference, MS-CG, and gGM $D_0$ values are from Table D1 in Appendix D, whereas the corrected $D_0$ values for the many-body Gaussian representation models were computed using Eq. (58).

| System | | | *Hard liquid* | | *Soft liquid* |
|---|---|---|---|---|---|
| | | | **Methanol** | **Acetonitrile** | **Acetone** |
| *Reference $D_0$ at the CG resolution* | | | 0.3866 | 0.3941 | 0.1705 |
| Corrected $D_0$ for CG models | *Actual model* | *MS-CG* | 0.3616 | 0.3390 | 0.2096 |
| | *Reduced model* | gGM | 0.4628 | 0.3816 | 0.1081 |
| | | MB-GRM | 0.4636 | 0.3579 | 0.2024 |

## V. Conclusions

The present work builds upon the previous papers in this series[61-64] to understand the FG and corresponding CG dynamics on the basis of the excess entropy scaling relationship. As the fifth paper of such series, the present work particularly focuses on a generalization of our earlier findings from the hard sphere system. Given quasi-universal scaling that holds for both FG and CG dynamics, the second paper of the current series was devoted to deriving the entropy-free diffusion coefficient term shown in excess entropy scaling to understand the fast CG dynamics in comparison with the reference dynamics at the FG resolution. However, this earlier work was established upon an assumption that the CG system can be approximated as a hard sphere. Due to a variety of molecular interactions and geometries, an extension to the non-hard sphere systems is necessary and inevitable.

To extend the hard sphere framework to non-hard sphere systems, we chose three different CG liquids that exhibit non-hard sphere characteristics: methanol, acetonitrile, and acetone. For these systems, we found that an additional repulsive interaction other than a hard sphere-like repulsion gives rise to much softer repulsion profiles, which is a signature of the non-hard sphere system. Therefore, we noted that naïvely applying the existing framework that was developed previously would fail to describe the correct CG dynamics. To resolve this limitation, we extracted the hard sphere characteristics from the non-hard sphere systems by looking at the signs of effective forces involving the local ordering based on the Weeks-Chandler-Andersen perturbation theory. Then, we scaled CG interactions to be effectively modeled as hard sphere. This approach can be readily applied for liquids as long as their inner-core repulsion resembles the hard sphere repulsion ("hard liquids"). For other types of liquids where the inner-core interactions decay much slower ("soft liquids"), we showed that the effective hard sphere nature can still be captured by examining density fluctuations at long wavelengths. To determine the changes in dynamics under this treatment, we derived the analytical correction factor term by considering pairwise contributions only. In turn, the corrected entropy-free diffusion coefficients are in a fair agreement with the reference data from the scaling relationship for both hard and soft liquids.

Taking a step further, we aimed to generalize the correction factor term by developing a simplified and generalized description for CG liquids. Specifically, we introduced *ad hoc* interaction models, i.e., CSW and GDW models, that have been used to model liquids to extend the applicability of the devised framework. Since these models are inherently simplified and not derived from the microscopic description of liquids, we extended our analysis by employing the many-body Gaussian representation to represent bottom-up CG interactions. Furthermore, we demonstrated that both simplified and bottom-up models are capable of reproducing interaction profiles while correctly recapitulating the entropy-free diffusion coefficients.

Altogether, this work extends an earlier treatment of the CG dynamics to understand an enhanced diffusion of more general systems at the CG level by leveraging excess entropy. Entropy in molecular coarse-graining has been understood both intuitively, as missing degrees of freedom, and rigorously, through information-theoretical approaches for deriving many-body CG PMFs (e.g., relative entropy[30-32] and relative resolution methods[205-207]). This paper series emphasize entropy's crucial role in coarse-grained dynamics, providing deeper insight into its application across various molecular systems. As such, one important direction from this work would be to elucidate the CG dynamics of large-scale systems including biomolecules and material systems. In such complex systems, nonnegligible contributions from the long-range interactions are more important than the hard sphere-like repulsion at short distances. Thus, it would be of great interest to rigorously extract the hard sphere nature from complex systems and determine the correction factor by including beyond pairwise contributions. We believe this work can facilitate the use of transferable CG modeling for bio-related systems, where reproducing correct kinetic information is a primary importance.

**ACKNOWLEDGMENTS**
This material is based upon work supported by the National Science Foundation (NSF grant CHE-2102677). Simulations were performed using computing resources provided by the University of Chicago Research Computing Center (RCC). The authors acknowledge insightful discussion with Dr. Yining Han and Professor Kenneth S. Schweizer. J.J. thanks the Arnold O. Beckman Postdoctoral Fellowship (http://dx.doi.org/10.13039/100000997) for funding and academic support.

**DATA AVAILABILITY**
The data that support the findings of this work are available from the corresponding author upon request.

**APPENDICES**

**A. Naïve Estimation of $D_0$ using the Carnahan-Starling EOS**

From Paper II[62] and Eq. (16b), the Carnahan-Starling EOS result in the following $D_0$ expression

$$D_{0,\mathrm{CS}}^{HS} \approx \frac{\pi^{\frac{1}{6}}}{48} \cdot 6^{\frac{4}{3}} \cdot \frac{(1-\eta)^3}{\eta^{\frac{2}{3}}(2-\eta)} \exp\left[\frac{(4\eta-3\eta^2)}{(1-\eta)^2}\right]. \qquad (A1)$$

Using the same calculation protocol described in Table 2, the results by employing Eq. (A1) are listed in Table A1. Since $D_{0,\mathrm{min}}(\eta_{\mathrm{min}})$ for the Carnahan-Starling EOS is 0.659 from Table 1, $D_0(\sigma_{\mathrm{BH}}^{\mathrm{naïve}})$ values are still higher than the predicted minimum. Note that the reference $D_0$ values from the scaling relationships are even smaller than $D_{0,\mathrm{min}}(\eta_{\mathrm{min}})$. Also, the $D_0(\sigma_{\mathrm{BH}}^{\mathrm{naïve}})$ values shown in Table A1 are in a comparable magnitude in comparison to the Percus-Yevick values, confirming that a choice of EOS does not affect any reasonings made in the main text.

**Table A1.** Naïvely estimated hard sphere diameter $\sigma_{\mathrm{BH}}^{\mathrm{naïve}}$ from the CG PMFs and the corresponding entropy-free diffusion coefficients $D_0$ ($\sigma_{\mathrm{BH}}^{\mathrm{naïve}}$) for soft and hard liquids. Note that the estimated $D_0$ ($\sigma_{\mathrm{BH}}^{\mathrm{naïve}}$) is larger than the minimum $D_0$ value derived from the Enskog kinetic theory using the Carnahan-Starling EOS $D_{0,\min}^{CS}(\eta_{\min})$. Since the naïve packing fraction for acetone significantly exceeds the physical limit, $D_0$ for acetone was not estimated.

| | *Hard liquid* | | *Soft liquid* |
|---|---|---|---|
| | **Methanol** | **Acetonitrile** | **Acetone** |
| $\sigma_{\mathrm{BH}}^{\mathrm{naïve}}$ | 3.828Å | 4.460Å | 7.680Å |
| $D_0$ ($\sigma_{\mathrm{BH}}^{\mathrm{naïve}}$) | 1.796 | 3.949 | - |
| *Reference* $D_0$ | 0.3866 | 0.3941 | 0.1705 |
| $D_{0,\min}^{\mathrm{CS}}(\eta_{\min})$ | | 0.659 | |

In Table A1, it is apparent that the naïve use of the Barker-Henderson approach results in a huge deviation by a relative error of 344.6% for methanol and 761.5% for acetonitrile. One thing to note is the acetone system where naïvely applying Eq. (A1) to the unchanged $\sigma_{\mathrm{BH}}^{\mathrm{naïve}} = 7.680\text{Å}$ gives completely wrong entropy-free diffusion coefficient $D_0(\sigma_{\mathrm{BH}}^{\mathrm{naïve}}) = -2.666 \times 10^{-3}$. The negative $D_0$ value can be understood from Eq. (19b) and Fig. 1 such that once $\eta > 0.8139$, $D_0$ decreases. Nevertheless, this packing fraction is way beyond the freezing fraction and the negative diffusion value further confirms its breakdown.

**B. Fluctuation Matching: Solution**

From the definition of the isothermal compressibility, the dimensionless compressibility at the zero wave number limit can be written as

$$S(k=0)_{\mathrm{HS}}^{\mathrm{CS}} = \rho k_B T \cdot \left(-\frac{1}{V} \cdot \frac{\partial V}{\partial P}\right) = \frac{(1-\eta)^4}{\eta^4 - 4\eta^3 + 4\eta^2 + 4\eta + 1}. \qquad (B1)$$

Unlike the Percus-Yevick case, Eq. (B1) does not have an analytic solution, and thus, by numerically solving Eq. (B1) with the Newton-Raphson method we obtain

$$S(k=0)_{\mathrm{HS}}^{\mathrm{CS}} = 0.4812, \qquad (B2)$$

giving $\eta$ =0.09328.

Finally, Table B1 summarizes the key results using the Carnahan-Starling EOS. We did not include the corrected $D_0$ values from the generalized Gaussian Models, because the Gaussian Model is not affected by the choice of EOSs, and thus the results should be identical to Table 6.

**Table B1.** Corrected estimated hard sphere diameter from the CG PMFs and the corresponding entropy-free diffusion coefficients $D_0$ for soft and hard liquids based on the non-hard sphere nature: $\sigma_{\mathrm{non-HS}}$ for soft liquids are obtained from the Barker-Henderson diameter and $\eta_{\mathrm{non-HS}}$ for hard liquids are obtained from fluctuation matching. Resultant $D_0$ (non-HS) values are corrected by including the correction factor for changes in entropic contributions. Here, we used the Carnahan-Starling EOS for treating the hard sphere system.

| System | *Hard liquid* | | *Soft liquid* |
|---|---|---|---|
| | **Methanol** | **Acetonitrile** | **Acetone** |
| *Effective hard sphere measure* | $\sigma_{\mathrm{non-HS}}$ | | $\eta_{\mathrm{non-HS}}$ |
| | 3.257Å | 3.175Å | 0.09328 |
| $D_0$ (non-HS) | 0.7193 | 0.6610 | 0.7969 |
| Corrected $D_0$ (non-HS) | 0.3645 | 0.3408 | 0.2097 |
| *Reference* $D_0$ | 0.3866 | 0.3941 | 0.1705 |

By extracting the hard sphere character interactions, large errors noticed in Table A1 are reduced to 86.06% for methanol, 67.72% for acetonitrile, and 367.4% for acetone. Consequently, by incorporating the changes in entropic terms, these errors further decrease by an error of 6.063% for methanol, 13.52% for acetonitrile, and 22.99% for acetone. In turn, the same conclusion can be derived from the Carnahan-Starling EOS.

**C. Generalized Gaussian Model for Arbitrary CG Liquids**

By generalizing the In this light, Eq. (45) is further generalized to the *generalized Gaussian model* $U_{\mathrm{gGM}}(R)$

$$U_{\mathrm{gGM}}(R) = \left(\frac{R_0}{R}\right)^{\delta_0} + \sum_{K=1}^{N_K} U_{R_K} \exp\left[-\frac{(R-R_K)^2}{2\delta_{R_K}^2}\right], \qquad (C1)$$

where $(R_0/R)^{\delta_0}$ denotes the hard sphere-like repulsive interaction at the inner-core region, and $U_{R_K} \cdot \exp\left[-(R-R_K)^2/2\delta_{R_K}^2\right]$ describes the localized repulsion ($U_{R_K} > 0$) or attraction ($U_{R_K} < 0$) along the radial domain. In principle, a total number of the Gaussian basis sets $N_K$ is highly dependent on the particular structure of liquids, since $N_K$ can be understood as a total number of local orderings, or packings, seen from the pair distribution. Namely, $N_K$ can be approximated as a total number of $R^*$ that satisfies $[dU/dR]|_{R=R^*} = 0$. From Fig. 2, $N_K$ is two for methanol (both repulsion and attraction), one for both acetonitrile and acetone (only repulsion). We expect this extension imparts much more descriptive features to the model to capture various local orderings compared to the CSW and GDW models.

Since a single Gaussian function $U_{R_K} \cdot \exp\left[-(R-R_K)^2/2\delta_{R_K}^2\right]$ is convex in the radial domain, its summation is still convex, which guarantees the convergence over the solution space. However, including a polynomial hard-core interaction $(R_0/R)^{\delta_0}$ makes this optimization no longer a convex optimization problem. However, for the finite $N_K$ values, parametrizing Eq. (C1) is still feasible by imposing the separation of interaction: $R_0 < R_1 < \cdots < R_K < R_{K+1} < \cdots < R_{N_K}$.

**D. Corrected CG Dynamics for Generalized Gaussian Models**

Having a generalized description of CG liquids, we now derive the effective hard sphere correction factor for the generalized Gaussian models. As shown in previous examples, it is reasonable to assume that the hard sphere characteristics are encoded in the short-range regime until $[dU_{\mathrm{gGM}}(R)/dR]|_{R=R_{\mathrm{HS}}} = 0$. $R_{\mathrm{HS}}$ can be determined analytically from the derivative of $U_{\mathrm{gGM}}(R)$

$$\frac{dU_{\mathrm{gGM}}(R)}{dR} = -\frac{\delta_0}{R}\left(\frac{R_0}{R}\right)^{\delta_0} - \sum_{K=1}^{N_K} \frac{U_{R_K}}{\delta_{R_K}^2}(R-R_K) \cdot \exp\left[-\frac{(R-R_K)^2}{2\delta_{R_K}^2}\right]. \qquad (D1)$$

Approximately, by employing the separation of interaction, we assume that the only first Gaussian repulsion with $U_{R_1} > 0$ affects near $R \approx R_{\mathrm{HS}}$, giving

$$0 = -\frac{\delta_0}{R_{\mathrm{HS}}}\left(\frac{R_0}{R_{\mathrm{HS}}}\right)^{\delta_0} - \frac{U_{R_1}}{\delta_{R_1}^2}(R_{\mathrm{HS}}-R_1) \cdot \exp\left[-\frac{(R_{\mathrm{HS}}-R_1)^2}{2\delta_{R_1}^2}\right]. \qquad (D2)$$

From Figs. 10–12, we can straightforwardly observe $R_0 < R_{\mathrm{HS}} < R_1$

$$\frac{\delta_0}{R_{\mathrm{HS}}}\left(\frac{R_0}{R_{\mathrm{HS}}}\right)^{\delta_0} = \frac{U_{R_1}}{\delta_{R_1}^2}(R_1 - R_{\mathrm{HS}}) \cdot \exp\left[-\frac{(R_1 - R_{\mathrm{HS}})^2}{2\delta_{R_1}^2}\right]. \quad (D3)$$

Both sides of Eq. (D3) are positive, and thus one can determine $R_{\mathrm{HS}}$ by iteratively solving the following equation:

$$\delta_0\delta_{R_1}^2\frac{R_0^{\delta_0}}{U_{R_1}} = R_{\mathrm{HS}}^{\delta_0+1}(R_1 - R_{\mathrm{HS}}) \cdot \exp\left[-\frac{(R_1 - R_{\mathrm{HS}})^2}{2\delta_{R_1}^2}\right]. \quad (D4)$$

Note that the left-hand side of Eq. (D4) is constant, whereas the right-hand side is only a function of $R_{\mathrm{HS}}$. After obtaining the $R_{\mathrm{HS}}$ by solving Eq. (D4), the adjusted interaction has a form of

$$U_{\mathrm{HS-gGM}}(R) = \left(\frac{R_0}{R}\right)^{\delta_0} + \sum_{K=1}^{N_K} U_{R_K}\exp\left[-\frac{(R - R_K)^2}{2\delta_{R_K}^2}\right] - U_{gGM}(R_{\mathrm{HS}}). \quad (D5)$$

Note that this adjustment scales the short-range regime ($K = 0, 1$) by

$$\left(\frac{R_0}{R}\right)^{\delta_0} + U_{R_1} \cdot \exp\left[-\frac{(R - R_1)^2}{2\delta_{R_1}^2}\right] - \delta_0\delta_{R_1}^2\frac{R_0^{\delta_0}}{U_{R_1}}\frac{1}{R_{\mathrm{HS}}^{\delta_0+1}(R_1 - R_{\mathrm{HS}})}. \quad (D6)$$

Using Eqs. (D1)–(D6), we obtained the adjusted CG interactions from the Gaussian models as depicted in Fig. D1. By comparing Fig. D1 to Fig. 8, it should be noted that both shifted interactions are almost indistinguishable. From the generalized description, we also computed the Barker-Henderson diameter for the shifted interactions, resulting in $\sigma_{\mathrm{non-HS-gGM}} = 3.040$Å for methanol, 4.166Å for acetonitrile, and 6.285Å for acetone.

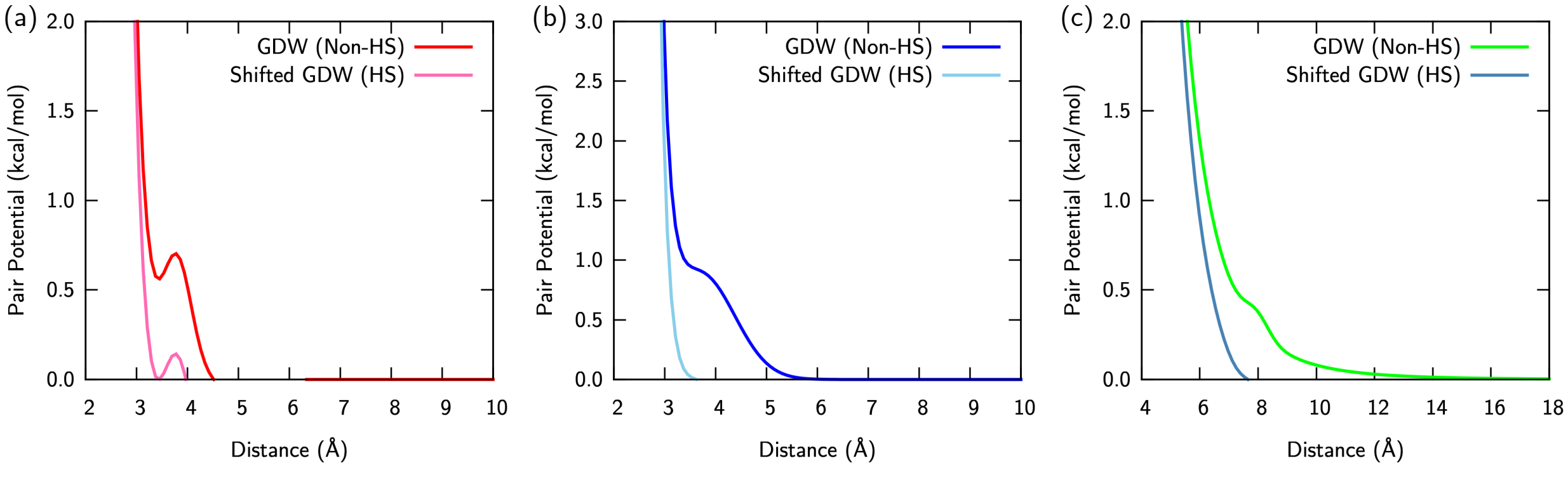

**Figure D1.** Non-hard sphere correction for the Gaussian liquid models of liquids based on the analytical solutions shown in Eq. (D5): (a) Methanol (b) Acetonitrile (c) Acetone. Note that the GDW models are plotted in red for methanol, blue for acetonitrile, and lime for acetone, and the shifted GDW models are shown in pink for methanol, sky blue for acetonitrile, and teal for acetone.

Recall that the “corrected” pairwise contribution to the excess entropy should be expressed as

$$S_2 = -2\pi\rho\int d\mathbf{R}\,[g(\mathbf{R}) \cdot (\ln g(\mathbf{R}) - 1) + 1]\mathbf{R}^2, \quad (D7)$$

which can be reformulated as follows for a spherically symmetric single-site CG system

$$S_2 \approx -2\pi\rho\int dR\,[e^{-\beta U_{\mathrm{HS-gGM}}(R)} \cdot \left(-\beta U_{\mathrm{HS-gGM}}(R) - 1\right) + 1]R^2. \quad (D8)$$

From the generalized CG Gaussian model, the hard sphere correction factor for any CG liquids can be determined as:

$$f \approx \frac{\exp\left(-\frac{2\pi\rho}{Nk}\int_0^\infty \left\{-\beta e^{-\beta U_{\mathrm{gGM}}(R)} U_{\mathrm{gGM}}(R) - \left[e^{-\beta U_{\mathrm{gGM}}(R)} - 1\right]\right\} R^2 \cdot dR\right)}{\exp\left(-\frac{2\pi\rho}{Nk}\int_0^\infty \left\{-\beta e^{-\beta U_{\mathrm{HS-gGM}}(R)} U_{\mathrm{HS-gGM}}(R) - \left[e^{-\beta U_{\mathrm{HS-gGM}}(R)} - 1\right]\right\} R^2 \cdot dR\right)}. \quad (D9)$$

To evaluate the accuracy of the Gaussian liquid model, we calculated $f$ and corrected the $D_0$ values.

**Table D1.** Generalized correction scheme for non-hard sphere systems using the Gaussian liquid model presented in this work. Inaccurately estimated naïve $D_0$ and the wrong $D_0$ (non-HS) are from Table 2. Finally, the corrected $D_0$ using the Gaussian model is presented in comparison to the non-parametrized MS-CG model, which is shown in Table 3.

| System | *Hard liquid* | | | | *Soft liquid* | |
|---|---|---|---|---|---|---|
| | **Methanol** | | **Acetonitrile** | | **Acetone** | |
| *Reference* $D_0$ | 0.3866 | | 0.3941 | | 0.1705 | |
| Wrong $D_0$ (hard sphere) | 1.845 | | 4.299 | | - | |
| Wrong $D_0$ (Non-hard sphere) | 0.7137 | | 0.6574 | | 0.7964 | |
| Corrected $D_0$ | *MS-CG* | *gGM* | *MS-CG* | *gGM* | *MS-CG* | *gGM* |
| (Non-hard sphere) | 0.3616 | 0.4628 | 0.3390 | 0.3816 | 0.2096 | 0.1087 |

By naïvely employing the Barker-Henderson approach regardless of the hard sphere nature of the system, a breakdown of the hard sphere model is observed by an error of 377.2% for methanol and 990.8% for acetonitrile. For acetone, this approach does not provide any realistic diffusion phenomena due to its soft inner-core nature. Witnessed from the non-hard sphere nature, the non-hard sphere PMFs can be adjusted to the hard sphere-like PMFs that can improve the corresponding $D_0$ by more than one order of magnitude: 84.61% for methanol, 66.81% for acetonitrile, and 368.0% for acetone. Nevertheless, these deviances from the reference data are due to a missing contribution from the excess entropy difference between the original and adjusted CG interactions. By carefully estimating the correction factor, we finally arrive at the correct entropy-free diffusion coefficient with a relative error of 6.467% for methanol, 13.98% for acetonitrile, and 23.17% for acetone.

Interestingly, despite its simplified nature, the generalized Gaussian models for these liquids can also demonstrate a fair reproduction of the $D_0$ values, indicating its applicability to other liquids in terms of recapitulating dynamics at the CG resolution. Given the extensive computational load required to parametrize the Gaussian representations in order to allow for transferability across different thermodynamic state points, a relatively simpler and cheaper parametrization of the generalized Gaussian models can be a desirable alternative to approximate the correction factor.